\documentclass{article}%%%

\usepackage{arxiv}

\usepackage[T1]{fontenc}    % use 8-bit T1 fonts
\usepackage{hyperref}       % hyperlinks
\usepackage{url}            % simple URL typesetting
\usepackage{booktabs}       % professional-quality tables
\usepackage{amsfonts}       % blackboard math symbols
\usepackage{nicefrac}       % compact symbols for 1/2, etc.
\usepackage{microtype}      % microtypography
\usepackage{graphicx}
\usepackage{doi}
\usepackage{array}
\usepackage{enumitem}
\usepackage{natbib}

\newcommand{\eps}{{\epsilon}}

\newcommand{\sig}{{\sigma}}

\newcommand{\Ome}{{\Omega}}
\graphicspath{{./fig/}}

\usepackage{amsmath} 
\usepackage{amssymb}
\usepackage{bm} 

\usepackage{mathtools}
\mathtoolsset{showonlyrefs}

\renewcommand{\u}{u}
\renewcommand{\d}{d}

\renewcommand{\eps}{\varepsilon}
\newcommand{\dmid}{\bar{d}}

\newcommand{\lc}{\ell_\mathrm{c}}
\newcommand{\lch}{\ell_\mathrm{ch}}
\newcommand{\sigc}{\sig_{c}}
\newcommand{\Uc}{U_\mathrm{c}}
\newcommand{\wc}{w_{c}}

\newcommand{\Gc}{G_\mathrm{c}}

\newcommand{\Ucal}{\mathcal{U}}
\newcommand{\Dcal}{\mathcal{D}}

\newcommand{\darg}{d'}
\newcommand{\uarg}{u'}
\newcommand{\harg}{h'}

\newcommand{\F}{F}
\newcommand{\FP}{\F_{\mathrm{P}}}
\newcommand{\FL}{\F_{\mathrm{L}}}
\newcommand{\Ffive}{F_5}
\newcommand{\Kfive}{K_5}
\newcommand{\Finf}{F_\infty}
\newcommand{\D}{\mathcal{D}}
\newcommand{\DP}{\D_{\mathrm{P}}}
\newcommand{\DL}{\D_{\mathrm{L}}}
\newcommand{\Dh}{\D^h}
\newcommand{\DPh}{\DP^h}
\newcommand{\DLh}{\DL^h}
\newcommand{\Ih}{\mathcal{I}^h}
\newcommand{\Lh}{\mathcal{L}^h}
\newcommand{\Sh}{\mathcal{S}^h}
\newcommand{\cP}{c_{\mathrm{P}}}
\newcommand{\cL}{c_{\mathrm{L}}}
\renewcommand{\r}{r}
\newcommand{\rP}{\r_{\mathrm{P}}}
\newcommand{\rL}{\r_{\mathrm{L}}}
\newcommand{\alphaP}{\alpha_{\mathrm{P}}}
\newcommand{\alphaL}{\alpha_{\mathrm{L}}}
\newcommand{\omegaP}{\omega_{\mathrm{P}}}
\newcommand{\omegaL}{\omega_{\mathrm{L}}}

\newcommand{\uP}{\u_{\mathrm{P}}}
\newcommand{\uL}{\u_{\mathrm{L}}}
\newcommand{\dP}{\d_{\mathrm{P}}}
\newcommand{\dL}{\d_{\mathrm{L}}}
\newcommand{\sigmaP}{\sigma_{\mathrm{P}}}
\newcommand{\sigmaL}{\sigma_{\mathrm{L}}}

\newcommand{\dint}{\mbox{\,  \ensuremath{\text{d}}}}

\newcommand{\nelt}{n}
\newcommand{\Ldem}{\frac{L}{2}}
\newcommand{\Udem}{\frac{U}{2}}
\newcommand{\nc}{\hat{i}}

\title{Phase-field and lip-field approaches for fracture with extreme mesh deformation (X-Mesh): a one-dimensional study}

\author{
 Nicolas Mo\"es \\
 Universit\'e catholique de Louvain \\
  Louvain-La-Neuve, Belgique \\
 \texttt{nicolas.moes.remacle@uclouvain.be} \\
   \And
Beno\^it L\'e \\
Ecole Centrale de Nantes\\
1 rue de La No\"e\\
 44321 Nantes, France \\
\texttt{benoit.le@cec-nantes.fr} \\
   \And
Nicolas Chevaugeon \\
Nantes Universit\'e \\
2, rue de la Houssini\`ere \\
 44322, Nantes cedex 3, France \\
\texttt{nicolas.chevaugeon@univ-nantes.fr} \\
\And
 Jean-Fran\c cois Remacle \\
 Universit\'e catholique de Louvain \\
  Louvain-La-Neuve, Belgique \\
 \texttt{jean-francois.remacle@uclouvain.be} \\
}

\renewcommand{\shorttitle}{\textit{Preprint submitted to} Comptes Rendus M\'ecanique}
\hypersetup{
pdftitle={Phase-field and lip-field approaches for fracture with extreme mesh deformation (X-Mesh): a one-dimensional study},
pdfauthor={Nicolas Mo\"es, Beno\^it L\'e, Nicolas Chevaugeon, Jean-Fran\c cois Remacle},
pdfkeywords={Phase-field, sharp cracks},
}

\begin{document}
\maketitle

\begin{abstract}
We consider a one-dimensional fracture problem modelled using either
the phase-field or lip-field approach.
In both cases, we optimise the incremental potential with
respect to the displacement and damage fields and the
nodal coordinates of the mesh.
This is thus a variational mesh study.
We observe that, as the damage reaches its maximum value, the optimisation drives the most damaged element
to zero size as the damage reaches its maximum value.
This peculiar element provides a precise displacement jump
representation as the bar breaks. The overall solution
is also shown to be much more accurate than the fixed mesh
solution. This work forms part of an exploration
into the capabilities of extreme meshes in computational mechanics (X-Mesh).
\end{abstract}


\section{Introduction}

The phase-field model as outlined by \cite{Bourdin2000,Karma2001,Bourdin2008,Miehe2010} has become a popular tool for studying material failure in recent years. It is based on minimising an incremental potential that depends on damage and displacement fields.
One advantage of this model is that the crack path does not need to be known in advance.
Cracks appear automatically during simulation and are located along the path where the damage $d$ has reached its maximum value (defined as $d = 1$ in this paper).
Another similar fracture model is the lip-field model \cite{Moes2021,Chevaugeon2022,Moes2022}. This model differs from the phase-field model in two ways.  Firstly, the objective incremental potential does not involve the damage gradient, making it equivalent to a non-regularised potential. Secondly, a constraint is added that requires the damage field to be Lipschitz. The use of a Lipschitz constraint on the damage field can be traced back to \cite{DalMaso2013}. Finally, note that the lip-field model is similar to the graded damage model introduced by \cite{Valoroso2022}.

The above regularised models often overlook the fact that the displacement field is discontinuous across the crack path.
This can be clearly observed, for instance, in analytical one-dimensional solutions of the phase-field model.
In this paper, we plan to investigate, among other things, whether this property is retained when a discrete model is considered, either with a fixed or an optimised mesh. By optimised meshes, we mean those based on the variational r-adaptation introduced in  \cite{mcneice1973,felippa1976,felippa1977,kuhl2004,askes2004,mosler2006,zielonka2008,Scherer2008,lahiri2010,munoz2017,tyranowski2019,shi2020}.Variational r-adaptation is closely linked to configurational forces, as discussed in \cite{braun1997,braun2007,steinmann2009,maugin2013,schmitz2023}. In the context of mesh optimisation, these forces vanish at the optimal node positions. These concepts have been applied to discontinuous fracture models, such as 'Griffith-like' models \cite{miehe2007,Scherer2007,Qinami2019} or cohesive models \cite{geissler2010,demaio2024}. Note that mesh optimisation based strategy have also been used to improve the accurary of phase-field results \cite{li2019,freddi2023}, but not in a variational framework (h-adaptativity).

Many models have attempted to incorporate discontinuities within continuous models using phase-field methods  \cite{Giovanardi2017,Geelen2018,Muixi2021,Zhang2022}, the Thick Level Set approach \cite{Moes2011,Bernard2012,Mororo2022} or other non-local damage models \cite{Seabra2013,Tamayo2014,Wang2016,Sarkar2021,Negi2022}. However, most of these models rely on specific discontinuous discretisation methods, such as X-FEM \cite{Moes1999}, and require the geometry of the macro-cracks to be identified from the continuous damage field. In this paper, we will demonstrate that displacement discontinuities can be naturally obtained by optimising the mesh using classical finite elements.

This work builds on that presented in \cite{Moes2023,Quiriny2024,Chemin2025}, which used extremely deformed meshes (X-Mesh) to track moving fronts. X-Mesh is a tracking approach with a twist. The twist is that it is not always the same nodes that carry the front. Some nodes join the front, while others leave it as it evolves. The mesh can become locally highly distorted, hence the name 'X-Mesh'. In the aforementioned works, X-Mesh was used for problems where the derivative of the main quantity of interest was discontinuous; for example, the temperature field in the Stefan phase-change problem \cite{Moes2023}. Since the classical finite element method only provides approximations that are continuous at the element boundaries, discontinuities can be captured by meshing the interface.

The paper is organised as follows. Section \ref{sec:model} details the one-dimensional phase-field and lip-field models chosen for the study.
The analytical solutions of both models are also presented.
Section \ref{sec:discretemodel}, presents the discrete models, and the finite element solutions obtained on fixed meshes are described.
Section \ref{sec:xmesh}, describes how the mesh is optimised, highlighting the improvement in the quality of the discrete solution. This is illustrated by some numerical results in section \ref{sec:results}. In particular, we analyse how and when the most damaged element becomes zero-sized. The paper concludes with a discussion and some thoughts for future work in section \ref{sec:conclusion}.

\section{Phase-field and lip-field models for the fracture of a bar}
\label{sec:model}

\begin{figure}
	\begin{center}
	\includegraphics[width=0.7\textwidth]{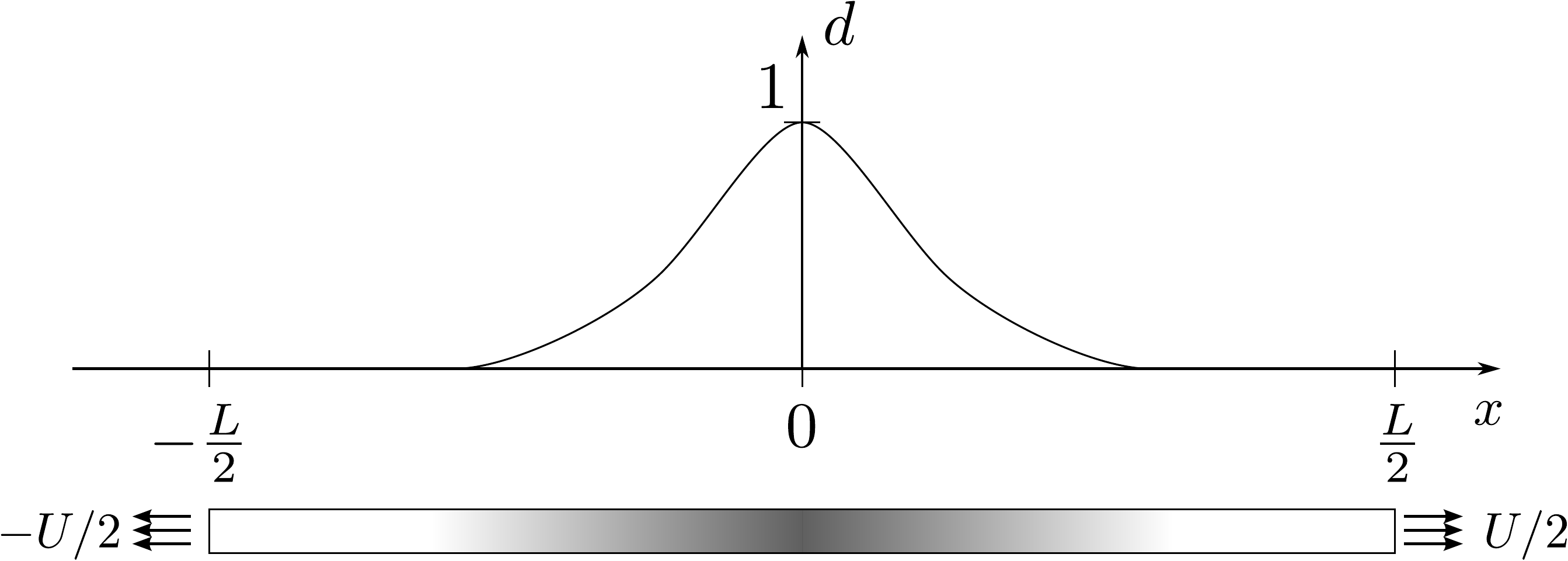}
	\end{center}
	\caption{A bar under tension with a symmetrical damage profile.}
	\label{fig:prob}
\end{figure}

\begin{figure}
	\begin{center}
\includegraphics[width=0.25\textwidth]{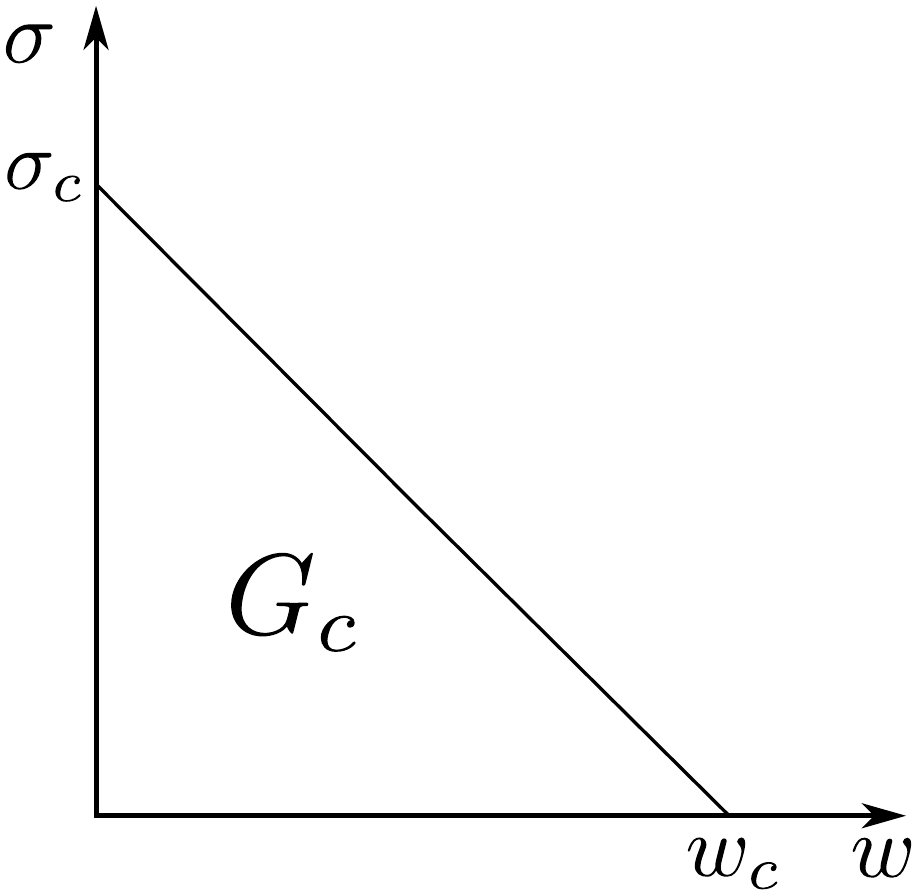}
	\end{center}
	\caption{Linear cohesive law. The area under the triangle corresponds to the toughness $\Gc \sigc \wc/2$.  As the opening  $w$ reaches the critical value $\wc$, the stress reaches zero.}
	\label{fig:cohesiveLinearLaw}
\end{figure}

\begin{figure}
	\begin{center}
\includegraphics[width=1.\textwidth]{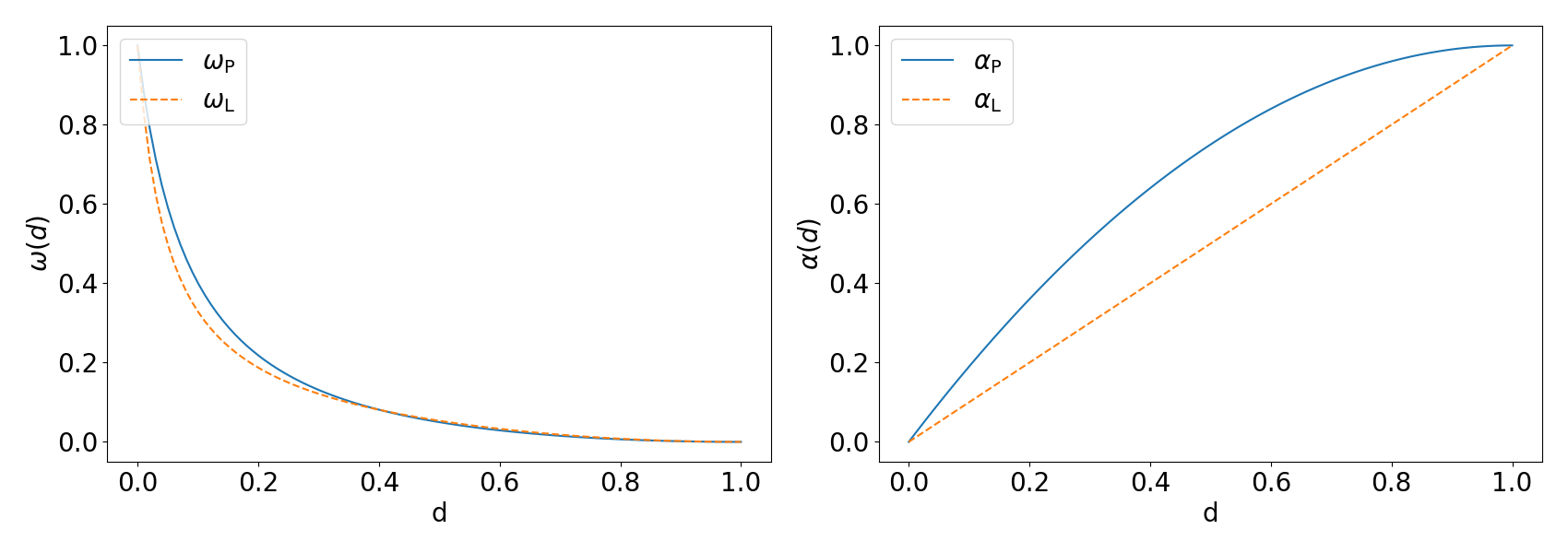}
	\end{center}
	\caption{Phase-field and lip-field $\alpha(d)$ and $\omega(d)$ functions, for $\gamma = 0.1$.}
\label{fig:phaseFieldModelFunctions}
\end{figure}

We consider the deformation of a one-dimensional bar, as depicted in Figure \ref{fig:prob} which occupies a one-dimensional domain $\Omega = \displaystyle \left[-\Ldem, +\Ldem \right]$.
A non-decreasing bar elongation $U(t)$, rising from zero,
is imposed on the bar in a symmetric fashion:
\begin{equation}
    u\left(\pm \Ldem,t \right) = \pm \frac{U(t)}{2}.
\end{equation}
The deformation of the bar at any point $x \in \Ome$ and any time $t$ is denoted by $\u(x,t)$ and the corresponding strain by $\eps(x,t)$.
Under the small strain assumption, the strain is the derivative of the deformation: $\eps = \u_{,x}$.
What follows considers a quasi-static evolution of $U(t)$, allowing the dependency on the time variable $t$ to be dropped and inertial effects to be neglected.
Material degradation is represented by a damage variable, $d \in [0,1]$; $d = 0$ corresponds to the undamaged material, $ 0 < d < 1$ to the damaged material and $d = 1$ to the fully damaged material. As material degradation is an irreversible process, $d$ cannot decrease over time.
The displacement field $\u$ and the damage field $d$ are solutions to a global minimisation problem
\begin{equation}
 (\u,d)  = \arg \min_{\uarg \in \Ucal, \darg \in \D} \F(\uarg, \darg) \label{eq:minimisationProblemOne}
\end{equation}
where $F$ is the incremental potential. The specific expression of $F$ depends on whether a phase-field or lip-field model is considered. We will use the subscripts ``$P$'' and ``$L$'' to distinguish between the two. If a subscript is not present, it means that the property is valid for both. It gives the following expressions of $\FP$ and $\FL$
\begin{align}
\FP(u,d) & = \int_{\Omega} \frac{1}{2}  \omegaP(d) E \eps^2(u) 
\dint x + \frac{\Gc}{\cP \lc} \int_{\Omega}  \alphaP(d) +   \lc^2 | \nabla d |^2  \dint x
\label{eq:phaseFieldFunctional}
\\
\FL(u,d) & = \int_{\Omega} \frac{1}{2}  \omegaL(d) E \eps^2(u) 
\dint x + \frac{\Gc}{\cL \lc} \int_{\Omega}  \alphaL(d)  \dint x 
\label{eq:lipFieldFunctional}.
\end{align}
The regularisation length is denoted $\lc$ and the scaling parameter $c$ is such that
 \begin{equation}
  c_P = 4 \int_0^1 \sqrt{\alpha_P(\beta)} \dint \beta, \quad c_L = 2 \int_0^1 \alpha_L(\beta) \dint \beta.
\end{equation}
The first term in the potential energy equation represents the stored elastic energy, whereas the second term represents the dissipated energy. In generic form, we will write:
\begin{equation}
F(u,d)  = \int_{\Omega} \frac{1}{2}  \omega(d) E \eps^2(u) 
\dint x + \frac{\Gc}{c \lc} \int_{\Omega}  \alpha(d) + r  \lc^2 | \nabla d |^2  \dint x
\end{equation}
where $\rP=1$ and $\rL=0$. The displacement field $\u$ belongs to the space of the kinematically admissible solutions
\begin{equation}
 \Ucal = \left\lbrace u \in H^1(\Omega) : u\left(-\Ldem \right) = -\frac{U}{2}, u\left(\Ldem\right) = \frac{U}{2} \right\rbrace
\end{equation}
whereas
\begin{align}
 \DP & = \{  d \in L^\infty(\Omega): \underline{d} \leq  d \leq 1     \} \label{eq:phaseFieldIrr} \\
 \DL & = \{  d \in L^\infty(\Omega): \underline{d} \leq  d \leq 1   \text{ and } | d(x)-d(y) | \leq | x-y | / \lc, \quad \forall x, y \in \Omega  \} \label{eq:lipIrr}
\end{align}
where $\underline{d}$ is the previously computed damage value. The spaces $\Ucal$ and $\Dcal$ detail the conditions on the displacement and damage fields, respectively. The degradation function $\omega(d)$ is such that
\begin{equation}
\omega(0) = 1, \quad \omega(1) = 0, \quad  \frac{\partial \omega}{\partial d}(d) \leq 0 \quad \forall d \in [0,1]
\end{equation}
whereas the dissipation function $\alpha(d)$ satisfies
\begin{equation}
\alpha(0) = 0, \quad \alpha(1) = 1,  \quad \frac{\partial \alpha}{\partial d}(d) \geq 0, \quad \forall d \in [0;1].
\end{equation}
Finally $E$ is the Young's modulus and $\Gc$ the material toughness. For the above models, the damage profile depends only on the maximum elongation reached so far. By imposing symmetry, the damage profile peaks at $x=0$ with the peak value denoted $d_0$. Since, we consider an increasing elongation starting from zero, we expect the response to first be elastic (no damage), then to show damage growth and finally to cease growing when $d_0$ reaches 1. Regarding the choice of the $\alpha$ and $\omega$ functions, we consider for $\alphaP$ and $\omegaP$ the model detailed in \cite{wu2017} for the phase-field and a new model for lip-field illustrated
 in Figure \ref{fig:phaseFieldModelFunctions}:
\begin{align}
\omegaP(d)   & = \frac{(1-d)^2}{(1-d)^2 + 2 \alphaP(d) / (\pi \gamma)},
  \quad 
  \alphaP(d) = 2 d -d^2
  ,  \quad \cP = \pi \\
    \omegaL(d) & = \frac{(1-d^2)^2}{(1-d^2)^2 + 2\alphaL(d)/\gamma}, \quad \quad \alphaL(d) = d, \quad \quad \quad \cL = 1
\end{align}
where we have used 
\begin{equation}
\label{eq:gamma}
    \gamma = \frac{\lc}{\lch}, \quad
    \lch = \frac{E \Gc}{\sigc^2}.
\end{equation}
 A value of $\gamma$ below $ \displaystyle \frac{8}{3 \pi}$ ensures the convexity of $\omegaP$ with respect to d, while the maximum value of $\gamma$ for $\omegaL$ is $ \displaystyle \frac{1}{2}$. Both are equivalent to a linear cohesive zone model (see Figure \ref{fig:cohesiveLinearLaw})
\begin{equation}
 \sigma = \sigc \left( 1 - \frac{w}{\wc} \right), \quad \wc = \frac{2 \Gc}{\sigc}.
\end{equation}
By ``equivalent'', we mean that the analytical solution to minimisation problem \eqref{eq:minimisationProblemOne} computed using the above functions or a linear cohesive zone model will produce the same elastic and dissipated energies. Usually, $\sigc$ and $\Gc$ are material parameters that can be obtained through experiments. Then, the cohesive equivalence is valid whatever the value of $\lc$, provided that the value of $\gamma$ in equation \eqref{eq:gamma} is small enough to ensure the convexity of $\omega$, which defines an upper bound on $\lc$.

\begin{figure}
	\begin{center}
	\begin{tabular}{c}
\includegraphics[width=0.6\textwidth]{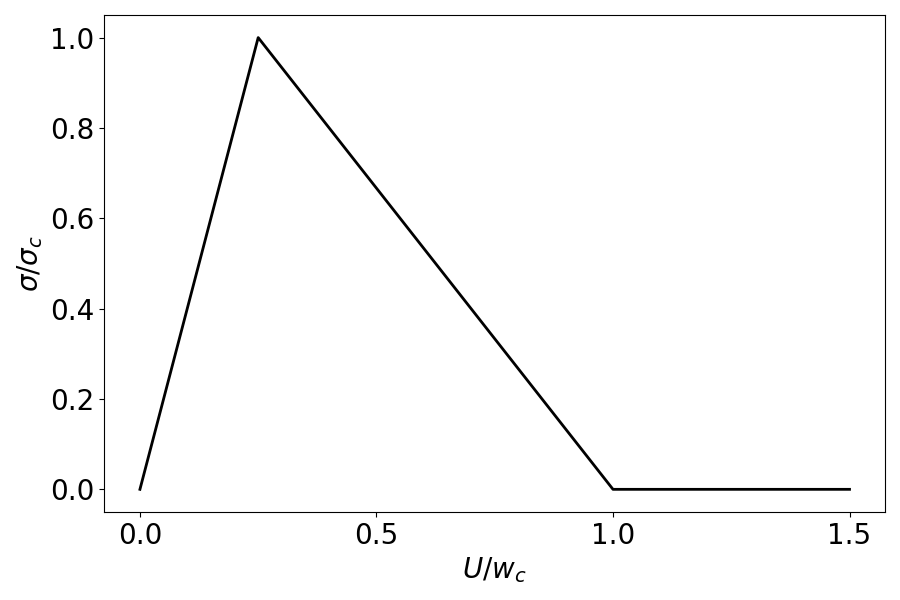}
\\ (a) Non-dimensional stress vs non-dimensioned imposed displacement.
\\
\includegraphics[width=0.8\textwidth]{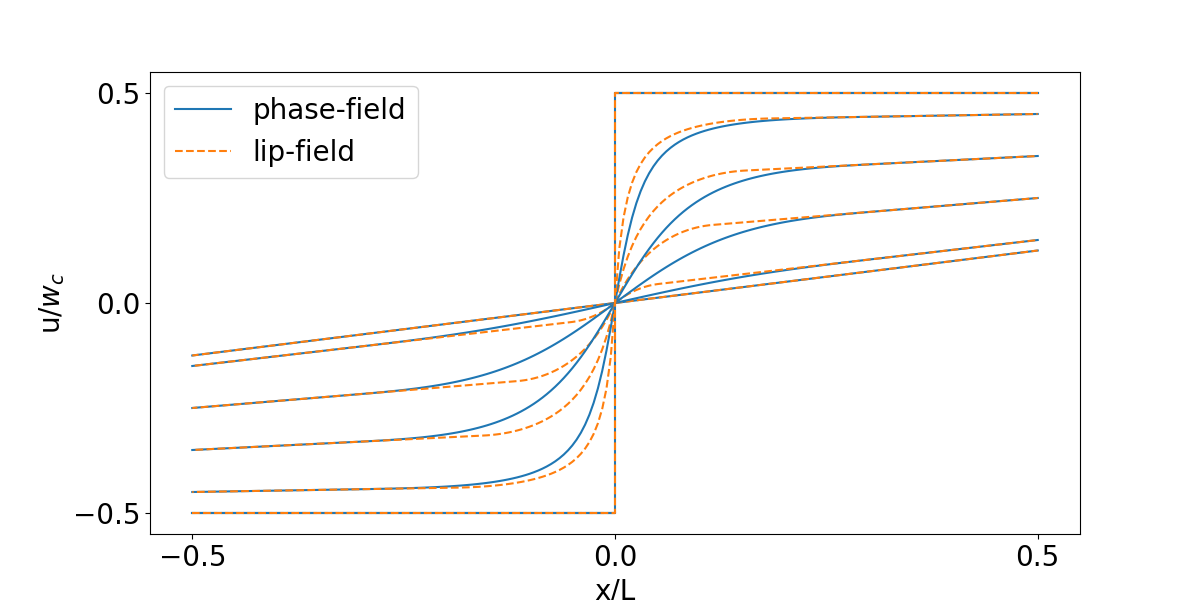}
\\ (b) Non-dimensional displacement.
\\
\includegraphics[width=0.8\textwidth]{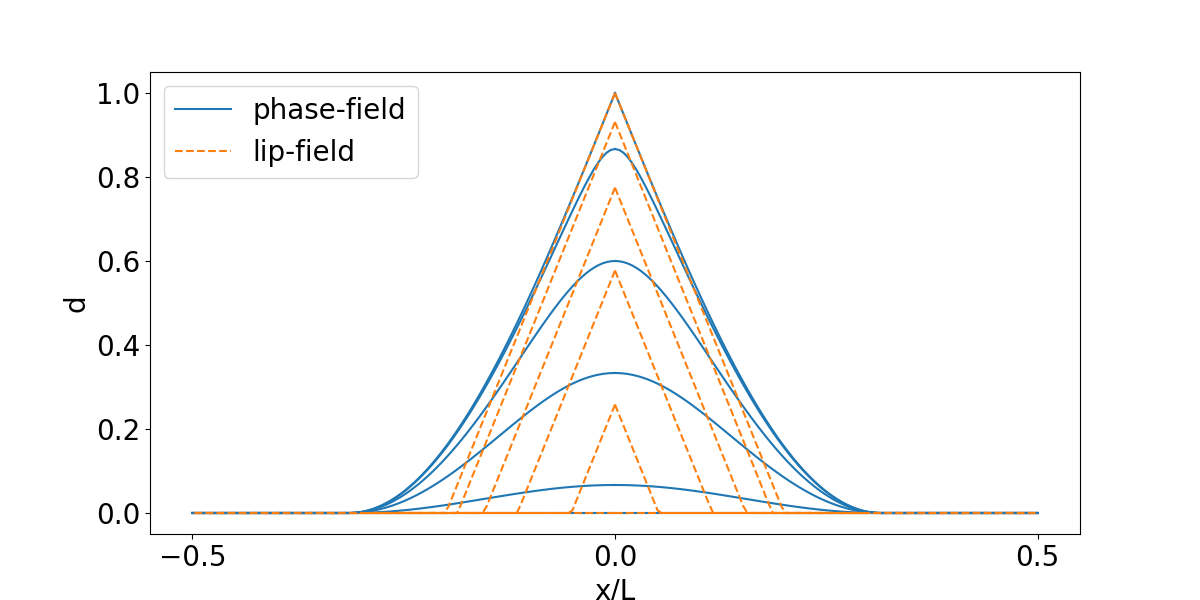}
\\ (c) Damage
	\end{tabular}

	\end{center}
	\caption{Analytical solution. The phase-field and lip-field models produce the same global response (a) but differ terms of displacement (b) and damage profiles (c) (the curves in Figures (b) and (c) are obtained for the same imposed displacement $U$).}
\label{fig:analyticalSolution}
\end{figure}

Figure \ref{fig:analyticalSolution} shows the analytical displacement and damage fields, as well as the force versus imposed displacement curves. The detailed expressions can be found in appendix \ref{appendix:phasefield} and \ref{appendix:lipfield}.The qualitative differences in the damage profiles between phase-field and lip-field are interesting to note (Figure \ref{fig:analyticalSolution} (c)). With phase-field  the width of the damaged zone is fixed, regardless of the imposed displacement $U$. In contrast, with lip-field the width of the damaged zone increases with $U$. Also, with phase-field, the damage profile is smooth at $x = 0$, except at bar breakage where it has a kink (with a slope of $\pm 1/\lc$).
The damage $d_0$ at the center of the bar begins at a critical stress denoted by $\sigc$, which corresponds to a bar elongation of $\Uc$.Then, damage gradually increases with elongation. Ultimately, damage reaches 1 at the critical elongation $\wc$:
\begin{equation}
	\label{eq:d0Pexact}
	d_0 = 
 \begin{cases}
     0 \quad \quad \quad \text{\ if\ } U \leq \Uc \\
     \frac{ U- \Uc}{\wc - \Uc } \quad \text{\ if\ } \Uc \leq  U \leq \wc \\
     1 \quad \quad \quad \text{\ if\ } U \geq \wc
 \end{cases}, 
\quad \Uc = L \frac{\sigc}{E}  
\end{equation}
while for lip-field,
\begin{equation}
	\label{eq:d0Lexact}
	d_0 =
 \begin{cases}
     0 \quad \quad \quad \text{\ if\ } U \leq \Uc \\
     \sqrt{\frac{ U- \Uc}{\wc - \Uc }} \quad \text{\ if\ } \Uc \leq  U \leq \wc \\
     1 \quad \quad \quad \text{\ if\ } U \geq \wc
 \end{cases},
\quad \Uc = L \frac{\sigc}{E}.
\end{equation}
For the models presented above to be exactly equivalent to a linear cohesive zone model, the entire damaged zone must lie strictly within the bar. At the same time, we will focus on cases with no snap-back, meaning damage must increase with an imposed increase in displacement. Combining these two conditions (equations \eqref{eq:d0Pexact} and \eqref{eq:PFDamagedBandWidth})) gives the following bounds for the bar length for phase-field
\begin{equation}
   \pi \lc \leq L \leq E \wc/\sigc
\end{equation}
which can be reformulated as
\begin{equation}
   \label{eq:barlPF}
    1 \leq \frac{L}{\pi \lc} \leq \frac{2}{\pi \gamma}
\end{equation}
while for lip-field,
\begin{equation}
 \label{eq:barlLip}
    1 \leq \frac{L}{\lc} \leq \frac{2}{\gamma}.
\end{equation}
For longer bars, a snap-back phenomenon occurs, whereby the damage evolution is discontinuous with respect to the
elongation. For shorter bars,
the expression is more complex because damage growth is active over the entire length of the bar.
Finally, in both models, the stress depends solely on the maximum damage value $d_0$
\begin{equation}
\label{eq:exactPFStress}
\sigmaP = \sigc (1-d_0)
\end{equation}
while for lip-field,
\begin{equation}
\label{eq:exactLipStress}
\sigmaL = \sigc (1-d_0^2).
\end{equation}

\begin{figure}
	\begin{center}
	\includegraphics[width=0.9\textwidth]{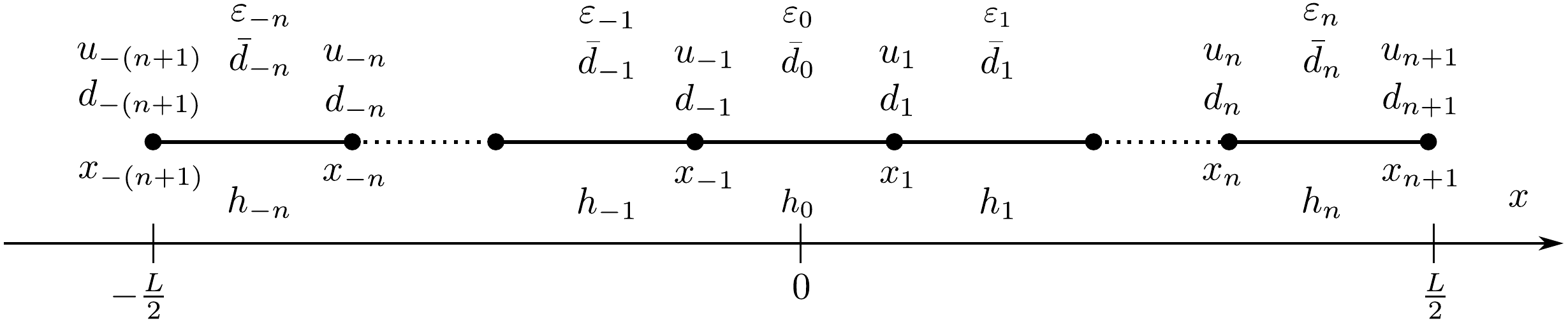}
	\end{center}
	\caption{Problem discretisation.}
	\label{fig:discr}
\end{figure}

\section{The discrete model}
\label{sec:discretemodel}

Figure \ref{fig:discr} illustrates the discretisation of the problem.
We consider an odd number of elements $N_e=2\nelt+1$. The central element has the ID 0.
The nodes are located at $x_i, i = \pm 1, \ldots, \pm(\nelt+1)$.
The set of nodal displacement and damage degrees of freedom are denoted by $u$ and
$d$, respectively. The set of element sizes is denoted by $h$.
\begin{equation}
    u = \{ u_i, i = \pm 1,  \ldots,  \pm(\nelt+1) \}, \quad d = \{ d_i, i = \pm 1,  \ldots,  \pm(\nelt+1) \}, \quad h = \{ h_i, i = 0, \pm 1, \ldots,  \pm \nelt \} \label{eq:range}.
\end{equation}
As stated in section \ref{sec:model}, we will consider a symmetric evolution of damage. Therefore, we can treat the above quantities as unknown for $i = 0 \ldots \nelt$, and then find their values for  $i = -1 \ldots -\nelt$ by symmetry.
Using a one-point integration scheme, the discrete potential is given by
\begin{align}
 F^h(u,d) & = \frac{1}{2} E \omega(d_0) \frac{(u_{1} - u_{-1})^2}{h_0} + \frac{\Gc}{c \lc}  h_0 \alpha(d_0) \\
 & + 2 \sum_{i = 1}^{ \nelt} \left( \frac{1}{2} E \omega(\dmid_i) \frac{(u_{i+1} - u_i)^2}{h_i} + \frac{\Gc}{c \lc} \left( h_i \alpha(\dmid_i)  + r \lc^2 \frac{(d_{i+1} - d_i)^2}{h_i} \right) \right)
\end{align}
where
\begin{equation}
\dmid_i =
\left\lbrace\begin{tabular}{l}
$ \displaystyle \frac{d_{-1} + d_{1}}{2} = d_0, \quad i = 0$ \\
$ \displaystyle \frac{d_{i} + d_{i+1}}{2}, \quad i =   1, \ldots,  \nelt. $
%  \\ $ \displaystyle \frac{d_{i} + d_{i-1}}{2}, \quad i =  - 1, \ldots, - \nelt $
\end{tabular}\right.
\end{equation}
The discrete potential must be minimised within the following constraints:
\begin{equation}
    u_{\nelt+1}  = \Udem, \quad 0 \leq d_i \leq 1, \quad h_i \geq 0, \quad h_0 + 2 \sum_{i = 1}^{ \nelt} h_i = L, \quad i = 0, \ldots,  \nelt.
\end{equation}
By optimising with respect to the displacement field, the stress is found to be uniform with a value of
\begin{equation}
\label{eq:sigDisc}
    \sigma = E K^h(d) \frac{U}{L},
%     \quad K^h(d) = L \left( \sum_{i = -\nelt}^{\nelt} h_i (\omega(\dmid_i))^{-1}   \right)^{-1}
    \quad K^h(d) = L \left(h_0 (\omega(d_0))^{-1} + 2 \sum_{i = 1}^{\nelt} h_i (\omega(\dmid_i))^{-1}   \right)^{-1}.
\end{equation}
We can then eliminate the displacement field in order to obtain
\begin{equation}
\label{eq:F}
 F^h(d,U) = F_e(U) K^h(d) + \Gc W^h(d)
\end{equation}
where 
\begin{equation}
 F_e(U) =  \frac{1}{2} \frac{E U}{L}^2
\end{equation}
and
\begin{equation}
 W^h(d) = \frac{h_0 \alpha(d_0)}{c \lc}  +
 \frac{2}{c \lc}  \sum_{i = 1}^{\nelt}
  \left( h_i \alpha(\dmid_i)  + r \lc^2 \frac{(d_{i+1} - d_i)^2}{h_i} \right).
\end{equation}
Regarding the constraints that $d$ and $h$ must satisfy, we define the following spaces:
\begin{align}
 \Sh & = \left\lbrace  (d,h) : \displaystyle h_0 + 2 \sum_{i=1}^{n} h_i = L \right\rbrace \label{eq:constSum} \\
 \Ih & = \left\lbrace (d,h) : \underline{d}(x_i) \leq  d_i \leq 1 \text{ and } \underline{d}_i \leq  d(\underline{x}_i)) \leq 1 , \quad i = 0 \ldots n \right\rbrace \label{eq:constIrr} \\
 \Lh & = \left\lbrace (d,h) : |d_{i+1}-d_i| \leq \lc h_i  , \quad i = 1 \ldots n \right\rbrace \label{eq:constLip}.
\end{align}
The first constraint \eqref{eq:constSum} imposes that the sum of the element lengths must be equal to the bar length. The second constraint \eqref{eq:constIrr} imposes irreversibility of the damage evolution. This amounts to requiring that the current nodal damage values $d_i$ are greater than the previous damage values interpolated at the nodes of the current mesh  $\underline{d}(x_i)$, and that the current damage values interpolated at the nodes of the previous mesh $d(\underline{x}_i)$ are greater than the previous nodal damage values $\underline{d}_i$. The third constraint \eqref{eq:constLip} is the Lipschitz constraint of the Lip-field model. The discrete versions of the spaces $\DP$ and $\DL$ introduced in equations \eqref{eq:phaseFieldIrr} and \eqref{eq:lipIrr} become
\begin{align}
  \DPh & = \Ih \\
  \DLh & = \Ih \cap \Lh.
\end{align}
Finally, the discrete problem is, for a given imposed displacement $U$,
\begin{equation}
 (d,h) = \arg \min_{(\darg,\harg) \in \Dh \cap \Sh} F^{\harg}(\darg,U).
\end{equation}
The problem is solved for $\{ i = 0, 1  \ldots,  \nelt \}$ using the \textit{scipy.optimize.minimize} Python routine \cite{Virtanen2020} \footnote{See also \textit{https://docs.scipy.org/doc/scipy/reference/generated/scipy.optimize.minimize.html\#scipy.optimize.minimize}}. Once $d$ and $h$ have been computed, $u$ can be obtained by solving an elasticity problem for which the damage is known.

\section{X-Mesh analysis}
\label{sec:xmesh}

The discrete solution of the one-dimensional fracture problem is obtained by minimising \eqref{eq:F} with respect to the damage $d$. In this section, we will analyse what happens when minimising with respect not only to $d$, but also with respect to the mesh element size $h$. Mesh optimisation consists in imposing
\begin{equation}
    \frac{\partial}{\partial h_i}
    \left( F^h + \lambda \left(h_0 + 2 \sum_{i = 1}^{\nelt}  h_i - L\right)\right) = 0, \quad i = 0, \ldots, \nelt
\end{equation}
where $\lambda$ is a Lagrange multiplier that enforces the condition $h_0 + 2 \sum_{i = 1}^{\nelt}  h_i = L$. This leads to
\begin{equation}
    - \frac{1}{2} E  \left( \frac{U}{L} \right)^2 (K^{h})^2 (\omega(\dmid_i))^{-1} + \frac{\Gc}{c \lc} \left( \alpha(\dmid_i) - r \lc^2 \left(\frac{d_{i+1} - d_i}{h_i}\right)^2 \right) + \lambda = 0.
\end{equation}
The Lagrange multiplier remains the same on each element. Provided at least one element remains undamaged, the value of $\lambda$ is
\begin{equation}
\label{eq:lambdaxmesh}
 \lambda = - (1/2) E (K^{h})^2) (U/L)^2
\end{equation}
In practice the length of the bar is such that the cohesive equivalence is valid, and therefore it is always possible to have an undamaged element (see end of section \ref{sec:model}). Furthermore, we force damage to localise on the central element using as a first damage guess in the optimisation process of the first loading step the analytical solution. This leads into
\begin{equation}
    \frac{1}{2} E \left( \frac{U}{L} \right)^2 (K^{h})^2 (1 - (\omega(\dmid_i))^{-1}) + \frac{\Gc}{c \lc} \left( \alpha(\dmid_i) - r \lc^2 \left(\frac{d_{i+1} - d_i}{h_i} \right)^2 \right) = 0
    \label{eq:hopti}.
\end{equation}
Applying the above to the central element, we have
\begin{equation}
    \frac{1}{2} E \left( \frac{U}{L} \right)^2 (K^{h})^2 (1 - (\omega(d_0))^{-1}) + \frac{\Gc}{c \lc} \alpha(d_0)   = 0
\end{equation}
or, using the stress expression, 
\begin{equation}
    \frac{1}{2} E^{-1} \sigma^2 (1 - (\omega(d_0))^{-1}) + \frac{\Gc}{c \lc} \alpha(d_0)   = 0
\end{equation}
giving for the phase-field model
\begin{equation}
\label{eq:sigPxmesh}
    \frac{\sigma}{\sigc} = (1 -d_0) 
\end{equation}
while the same computation gives for the lip-field model
\begin{equation}
\label{eq:sigLxmesh}
    \frac{\sigma}{\sigc} = (1 -d_0^2).
\end{equation}
The X-Mesh enables the exact stress-maximum damage relationship to be recovered for both models. This is an important result. Through mesh optimisation, we can see that the stress-damage relationship corresponds to the reference relationship, regardless of the value of $d_0$ or the number of elements (provided at least one element remains undamaged).
Note that if the equivalence with a linear cohesive zone model is not satisfied, for instance if we take a bar that is shorter than $2 \lc$, we cannot use equation \eqref{eq:lambdaxmesh}. Therefore all the equations up to \eqref{eq:sigPxmesh} and \eqref{eq:sigLxmesh} are invalid. This makes sense, since these two equations stems from the equivalence with the linear cohesive zone model.

Applying equation \eqref{eq:hopti} to any element other than the central one gives us the following for the phase field:\begin{align}
    \frac{\lc^2}{h_i^2} (d_{i+1} - d_i)^2  & = \alpha (\dmid_i) \left (1 - \left( \frac{1-d_0}{1-\dmid_i}\right)^2 \right) = (H( \dmid_i, d_0))^2, \quad 
    i = 1, \ldots, n-1.
\end{align}
The damage gradient over each element is thus related to the average
damage value of the element, as defined by the reference relation (see equation \eqref{eq:gradex}  in Appendix \ref{appendix:phasefield}).

Regarding lip-field, we will now prove that for X-Mesh, the damage gradient may only be 0 or $\pm 1/\lc$, and not any intermediate values as it is the case for a fixed mesh. We will prove this result for any element $i > 0$, since the $i = 0$ case, which corresponds to the central element, is trivial ($d_1 = d_{-1}$ by symmetry so the gradient is necessarily zero). Let us introduce the Lagrange multiplier $\mu_i$ which imposes the Lipschitz condition
\begin{equation}
\label{eq:lipLagrangeMultiplier}
 \mu_i \geq 0, \quad \frac{|d_i - d_{i+1}|}{h_i} - \frac{1}{\lc} \leq 0, \quad \mu_i \left( \frac{|d_i - d_{i+1}|}{h_i} - \frac{1}{\lc} \right) = 0, \quad i = 1 \ldots \nelt
\end{equation}
so mesh optimisation gives
\begin{equation}
    \frac{\partial}{\partial h_i}
    \left( F^h + \lambda \left(h_0 + 2 \sum_{i = 1}^{\nelt} h_i - L \right) + \sum_{1}^{\nelt} \mu_i \left( \frac{|d_i - d_{i+1}|}{h_i} - \frac{1}{\lc} \right) \right) = 0, \quad i = 1, \ldots, \nelt.
\end{equation}
\begin{equation}
    \frac{1}{2} E \left( \frac{U}{L} \right)^2 (K^{h})^2 (1 - (\omega(\dmid_i))^{-1}) + \frac{\Gc}{\lc}  \alpha(\dmid_i) - \frac{\mu_i |d_i - d_{i+1}|}{h_i^2} = 0, \quad i = 1, \ldots, \nelt
\end{equation}
which can be simplified using equations \eqref{eq:sigDisc} and \eqref{eq:sigLxmesh} to
\begin{equation}
\label{eq:lipOptCondition}
    \frac{\Gc}{\lc}  \alpha(\dmid_i) \left(1- \frac{(1-d_0^2)^2}{(1-\dmid_i^2)^2}  \right) - \frac{\mu_i |d_i - d_{i+1}|}{h_i^2} = 0, \quad i = 1, \ldots, \nelt.
\end{equation}
In what follows, we assume that $h_i > 0, \forall i >  0$, and demosntrate that the above equation  implies that the damage gradient is either 0 or $\pm 1/\lc$. We will distinguish between four cases:
\begin{itemize}
 \item $d_i = d_{i+1} = 0$ : then we have that $\dmid_i = 0$, so $\alpha(\dmid_i) = 0$. As the two terms in \eqref{eq:lipOptCondition} are null, this case is possible, corresponding to a fully elastic element where the gradient of $d$ is zero.
  \item $d_i = d_{i+1} = d_0$ : this gives  $\dmid_i = d_0$, so $1 - \displaystyle \frac{(1-d_0^2)^2}{(1-\dmid_i^2)^2} = 0$ The two terms in equation \eqref{eq:lipOptCondition} are null which makes this case possible. This could correspond to a homogeneous evolution of damage on the bar, for instance.
  \item $d_i = d_{i+1} \neq 0$ and $d_i = d_{i+1} \neq d_0$: Since $d_i = d_{i+1} \neq 0 $, we have $\alpha(\dmid_i) > 0$, and $d_i = d_{i+1} \neq d_0$ gives $1 - \displaystyle \frac{(1-d_0^2)^2}{(1-\dmid_i^2)^2} \neq 0$. The first term of \eqref{eq:lipOptCondition} is not zero, but the second term in \eqref{eq:lipOptCondition} is zero, which is not possible.
 \item $d_i \neq d_{i+1}, \quad d_i > 0 \text{ or } d_{i+1} > 0, \quad d_i < d_0 \text{ or } d_{i+1} < d_0 $. Since $d_0 > \dmid_i \quad \forall i >0$, then $1 - \displaystyle \frac{(1-d_0^2)^2}{(1-\dmid_i^2)^2} >0$ so the first term of the above equation is positive. Since $d_i \neq d_{i+1}$, $\mu_i$ must be strictly positive for the above equation to be zero. Combining this last condition with equation \eqref{eq:lipLagrangeMultiplier}, imposes that the gradient of the damage field must be $\pm 1/\lc$ on element $i$.
\end{itemize}

\section{Numerical experiments}
\label{sec:results}

The values of the different parameters for the numerical experiments are given in Table \ref{table:materialProperties}. Unless otherwise specified, the number of elements per width of the fully damaged zone (which is equal to $\frac{\pi \lc}{2}$ for phase-field and $\lc$ for lip-field) is $n_c = 5$. Analytical curves will be plotted in black, plain lines, while numerical results will be plotted in coloured, dotted lines.

\begin{table}
\centering
\begin{tabular}{llcc}
  \hline
  Properties  & Units     &  Symbol    &  Value   \\ \hline
  Bar length & m & $L$ & 0.2 \\
  Regularization length & m & $\lc$ & $L/5$ = 0.04 \\
  Young's modulus & Pa &  $E$    &  $ 3 \cdot 10^{10}$    \\
  Fracture toughness & N/m & $G_c$ & $120$ \\
  Critical tensile stress & Pa & $\sigc$ & $3 \cdot 10^6$ \\
  \hline
 \end{tabular}
 \caption{Dimensions an material properties for numerical example.}
 \label{table:materialProperties}
\end{table}

\subsection{Phase-field analysis}

\begin{figure}
\begin{center}
\includegraphics[width=9cm]{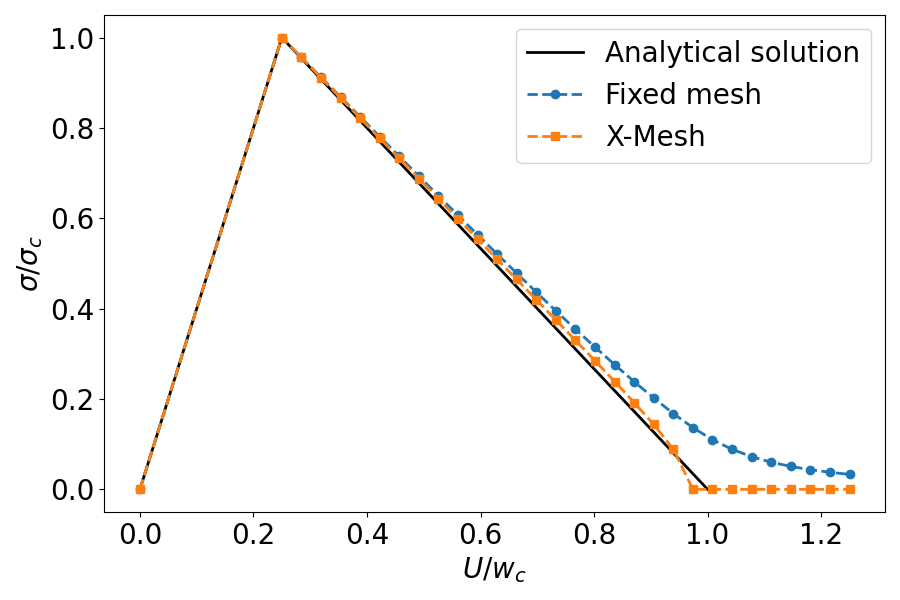}
\\ (a) Stress versus imposed displacement (phase-field)
\\
\begin{tabular}{cc}
\includegraphics[width=8cm]{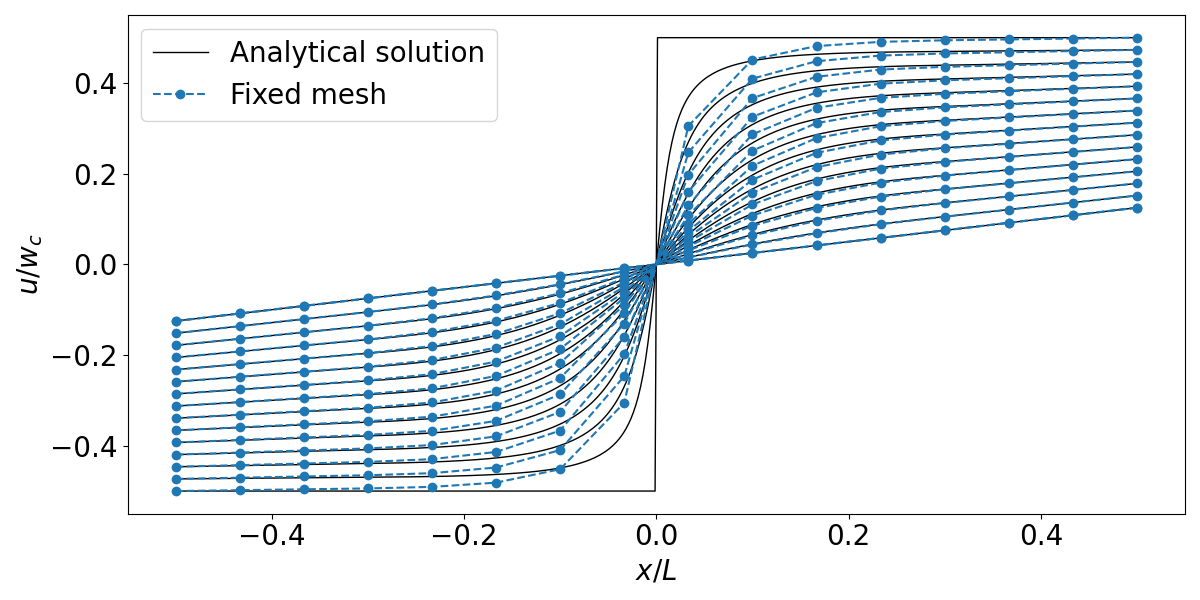}
& \includegraphics[width=8cm]{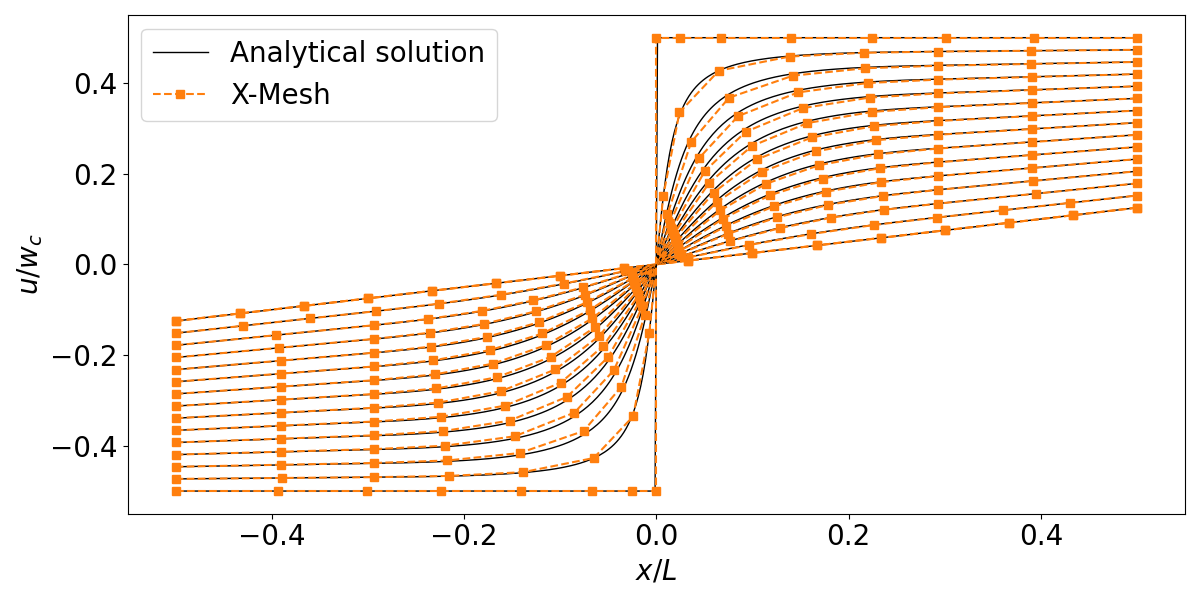}
\\
\end{tabular}
\\ (b) Displacement fields (phase-field)
\\
\begin{tabular}{cc}
\includegraphics[width=8cm]{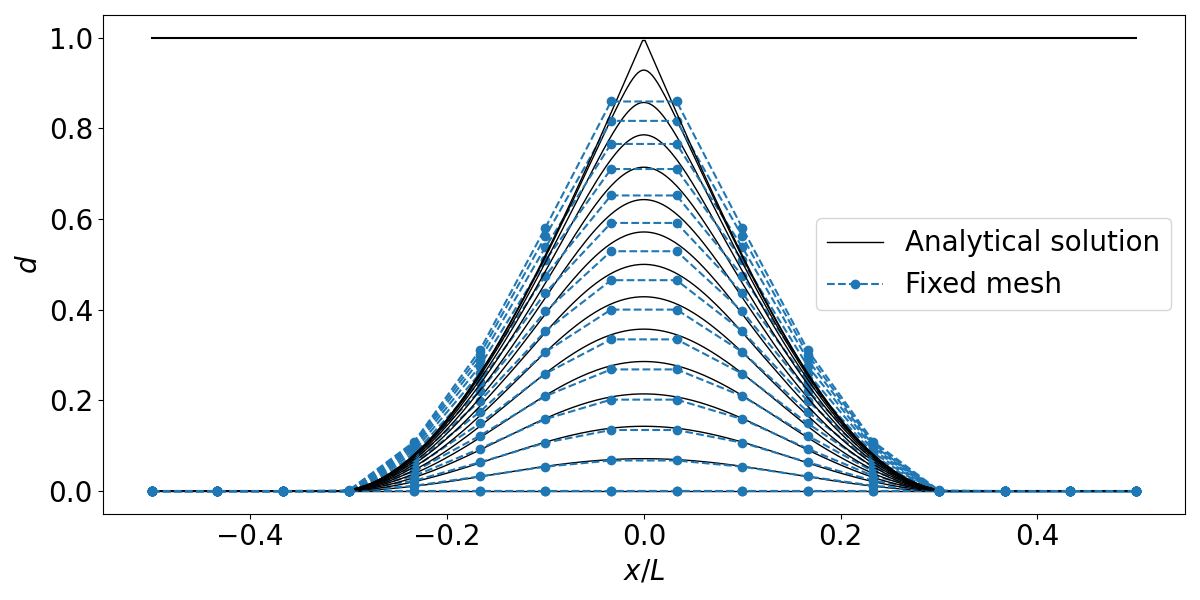}
& \includegraphics[width=8cm]{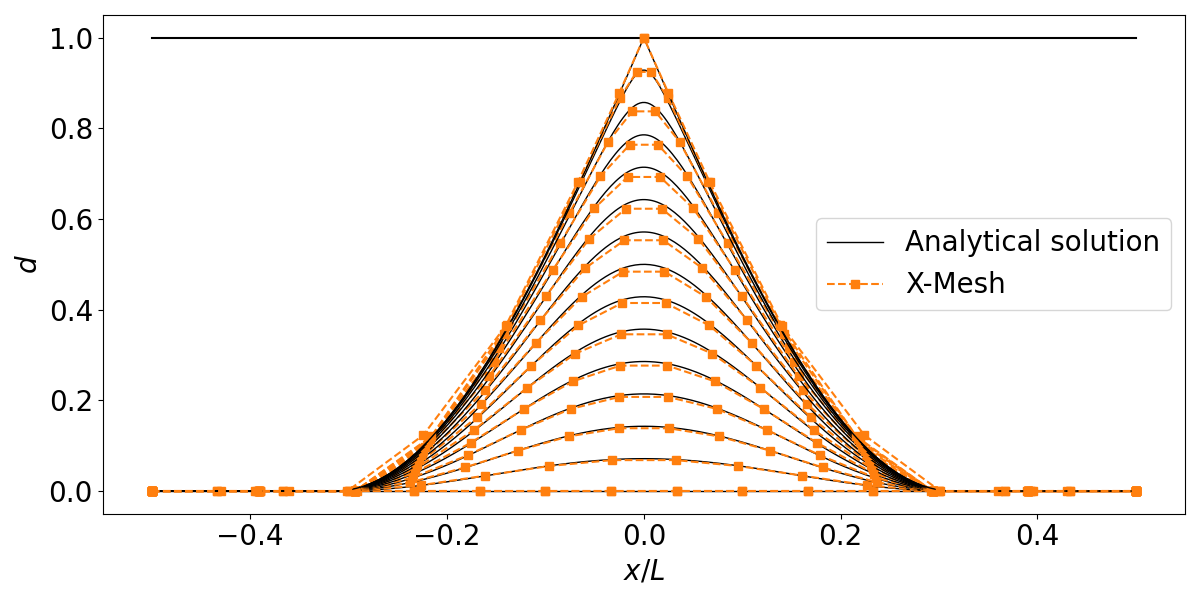}
\\
\end{tabular}
(c) Damage fields (phase-field)
\begin{tabular}{cc}
\includegraphics[height=5cm]{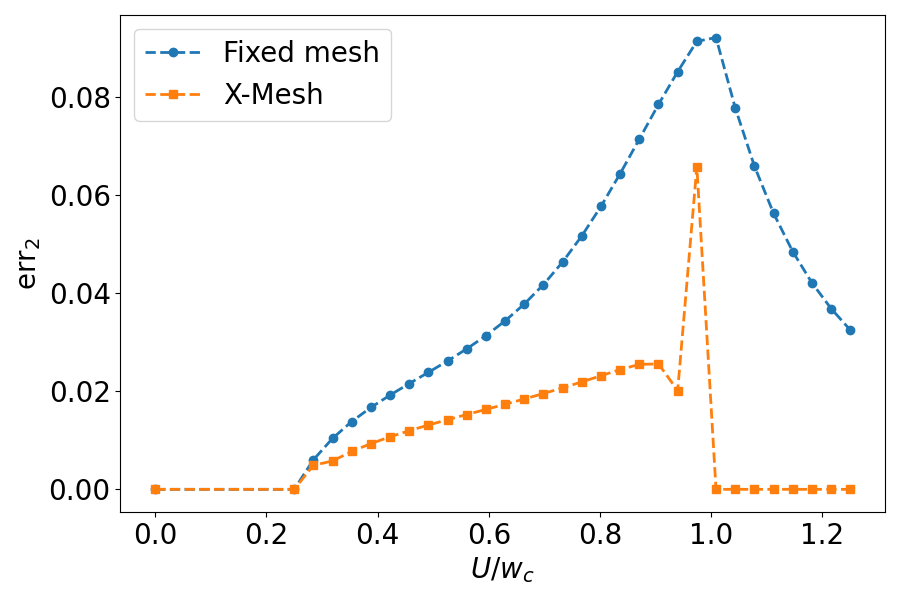}
& \includegraphics[height=5cm]{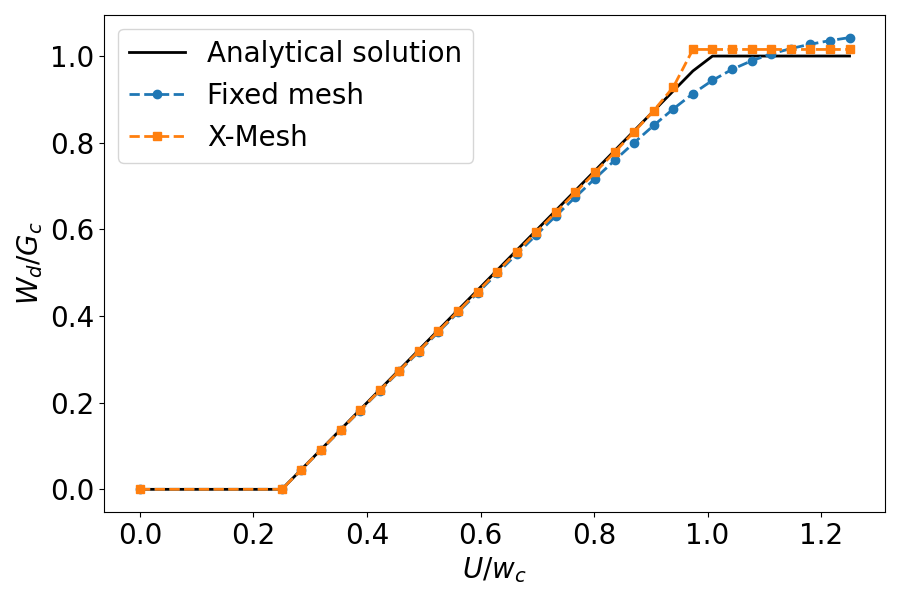}
\\
(d) $L_2$ error on displacement (phase-field) & (e) Dissipated energy (phase-field)
\end{tabular}
\end{center}
\caption{Phase-field: results obtained with a fixed mesh and X-Mesh, for five elements per $\frac{\pi}{2} \lc$.}
\label{fig:phaseField1DDispAndDamage}
\end{figure}

\begin{figure}
\begin{center}
\includegraphics[width=12cm]{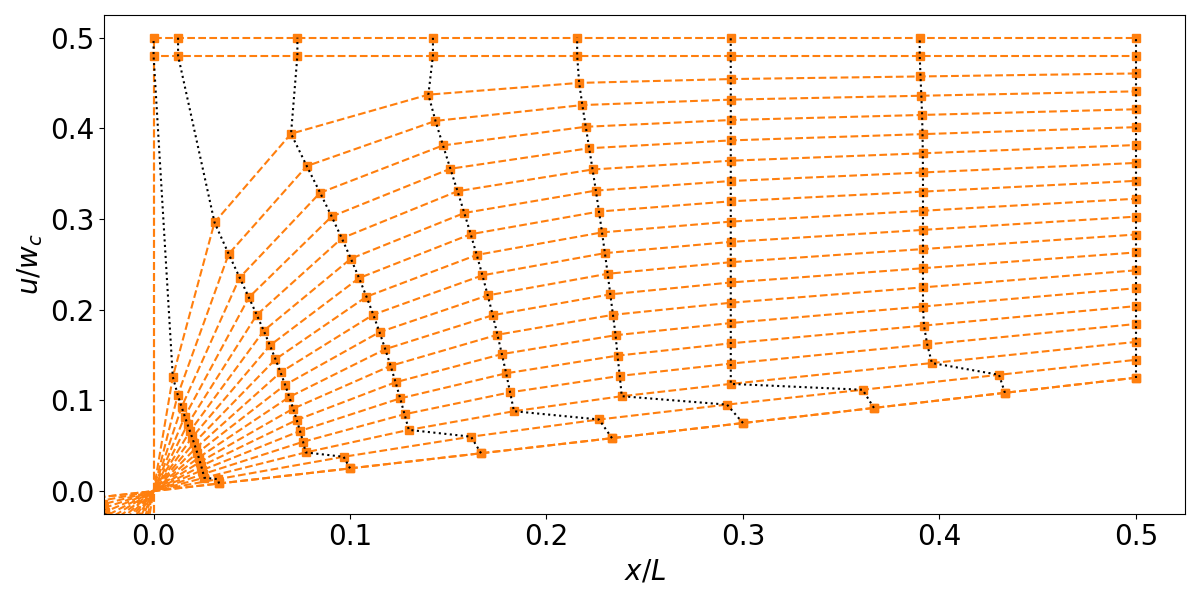}
\end{center}
\caption{Phase-field: Nodes displacement during X-Mesh computation.}
\label{fig:PhaseFieldXMeshNodesTrajectories}
\end{figure}

Some first results obtained using phase-field are presented in Figure \ref{fig:phaseField1DDispAndDamage}. The stress $\sigma$, computed using equation \eqref{eq:sigDisc}, is plotted as a function of the imposed displacement $U$ in Figure \ref{fig:phaseField1DDispAndDamage} (a) (Note that each dot corresponds to an increment of $U$). It shows that, with a fixed mesh, the bar never breaks; that is to say, the stress never reaches zero even at high values of $U$. Analysis of the discrete scheme shows that infinite elongation is required to break the bar. Conversely, with X-Mesh the bar breaks, albeit for an imposed displacement slightly smaller than the analytical value of $U = \wc$. The displacement fields are plotted in Figure \ref{fig:phaseField1DDispAndDamage} (b). With a fixed mesh, the displacement jump when $U = \wc$ cannot be captured. However for X-mesh, the size of the central element decreases to zero, enabling a displacement jump to be obtained. It can be seen that not only do the central nodes move, but the others do too, tending to move towards areas of higher displacement gradient.  Zooming in on Figure \ref{fig:PhaseFieldXMeshNodesTrajectories} emphasises this phenomenon. Figure \ref{fig:phaseField1DDispAndDamage} (c) shows the damage distribution on the bar. With X-Mesh, the central element has a damage value of 1, which is not the case with a fixed mesh. This explains why the bar breaks with X-Mesh but not with a fixed mesh.

Figure \ref{fig:phaseField1DDispAndDamage} (d) shows the $L^2$ error defined as
\begin{equation}
\label{eq:errL2}
 err_2 = \displaystyle \sqrt{\frac{\int_\Omega |\u-\u_{ex}|^2 \dint \Omega }{\int_\Omega |\u_{ex}|^2 \dint \Omega}}
\end{equation}
where $\u$ is the numerical solution of the problem and $\u_{ex}$ the analytical one (given by equation \eqref{eq:uExP} or \eqref{eq:uExL}). Using a middle point integration, it is computed as
\begin{equation}
\label{eq:errL2Disc}
 err_2^h = \displaystyle \sqrt{\frac{ \sum_{i=-\nelt}^{\nelt} h_i |\bar{\u}_i-\bar{\u}_{ex,i}|^2  }{\sum_{i=-\nelt}^{\nelt} h_i |\bar{\u}_{ex,i}|^2 }}
\end{equation}
where $\bar{\u}_i$ and $\bar{\u}_{ex,i}$ are the numerical and exact displacement fields evaluated at the middle of element $i$. As can be seen, for $U > \wc$, the error obtained with X-Mesh is exactly zero, confirming the results shown in Figure \ref{fig:phaseField1DDispAndDamage} (b). However, the error levels obtained with X-Mesh are not only lower when the bar is broken, but also for $U > \wc$. There is an error peak just before $U$ reaches $\wc$ which corresponds to the sudden stress drop shown in Figure \ref{fig:phaseField1DDispAndDamage} (a), which we will attempt to explain later.

Figure \ref{fig:phaseField1DDispAndDamage} (e) shows the dissipated energy as
\begin{equation}
\label{eq:dissipation}
 W_d(t) =   \int_0^t \sigma \dot{U} \dint t - \frac{1}{2}  \int_\Omega E \omega(d) \eps^2 \dint \Omega.
\end{equation}
The first term in the above equation represents the energy supplied to the system, while the second term represents the stored elastic energy. The discrete counterpart is computed using a midpoint rule for spatial integration and a trapezoidal rule for temporal integration
\begin{equation}
\label{eq:dissipationDisc}
 W_d^N =  \sum_{k=1}^{N}  \frac{\sigma^k+\sigma^{k-1}}{2} (U^k+U^{k-1}) - F_e(U^N) K^h(d^N).
\end{equation}
As an exception, in the above formula we noted quantities at the current time step with a $N$ exponent, and quantities at the previous time steps with a ``$k = 0 \ldots N-1$'' exponent.

Up to a certain imposed displacement the dissipated energy for the fixed mesh and for X-Mesh are the same. Then as $U$ gets close to $\wc$, the dissipated energy for the fixed mesh continues to increase and becomes higher than $\Gc$. This is in agreement with the damage numerical damage profile which is wider than the analytical one showwn in Figure \ref{fig:phaseField1DDispAndDamage} (c). It also seems to keep increasing even for values of $U$ greater than $\wc$. Conversely, with X-Mesh, the dissipated energy stabilises to a value which is much closer to $\Gc$.

\begin{figure}
\begin{center}
\begin{tabular}{cc}
\includegraphics[width=8cm]{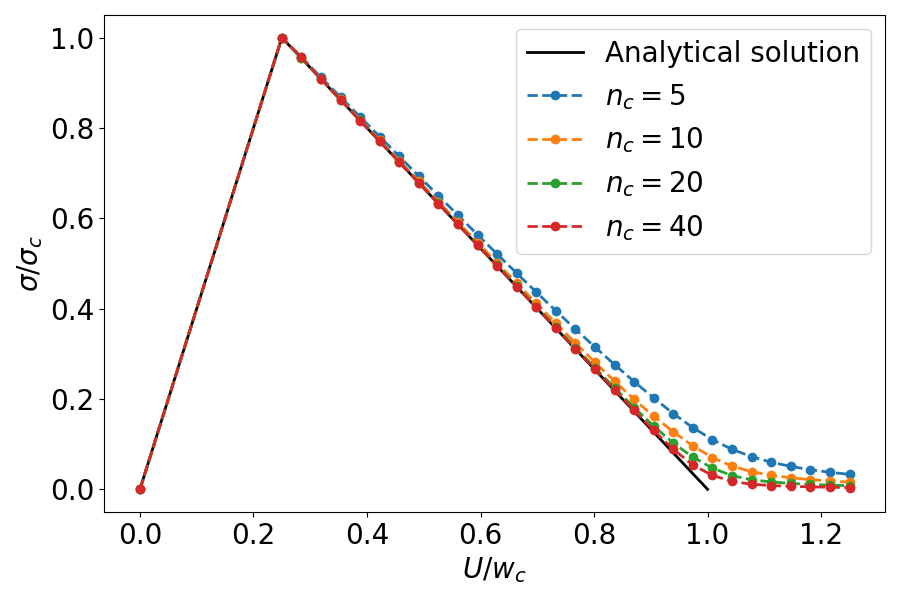}
&
\includegraphics[width=8cm]{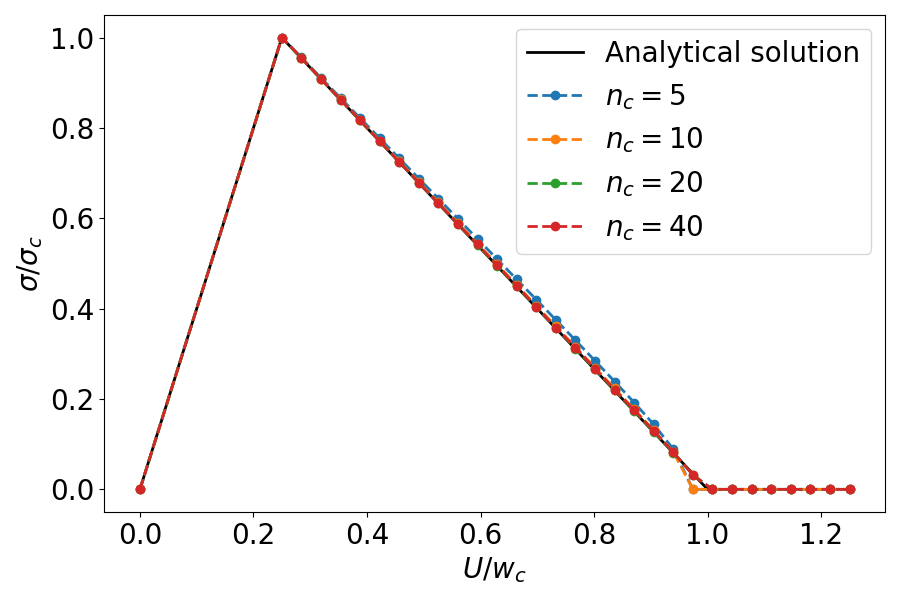}
\end{tabular}
\end{center}
\caption{Phase-field: Stress versus imposed displacement curves obtained with a fixed mesh (left) and for X-Mesh (right) for different values of $n_c$.}
\label{fig:PhaseFieldXMeshAndFixedMeshConvergenceSigU}
\end{figure}

\begin{figure}
\begin{center}
\begin{tabular}{cc}
% FEM & X-Mesh \\
\includegraphics[width=8cm]{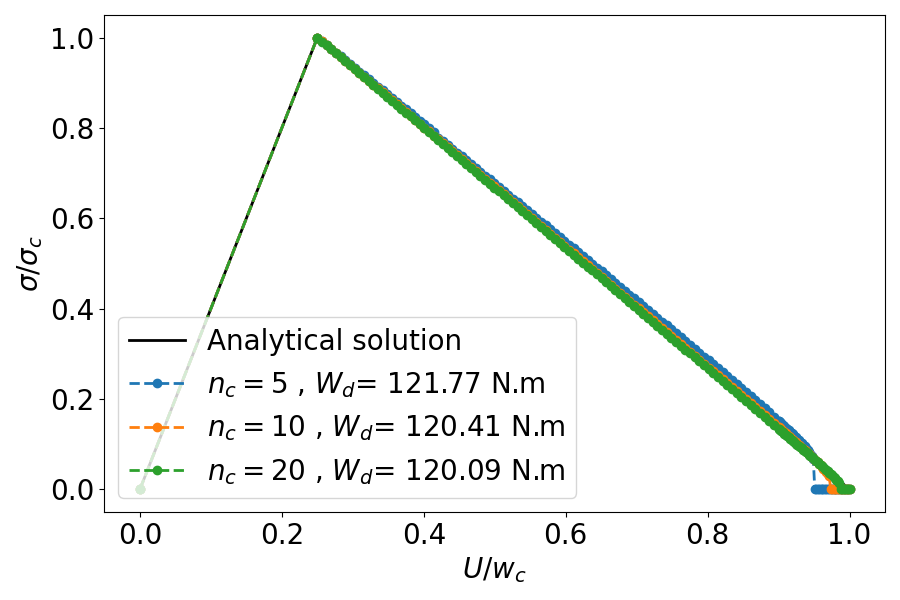}
&
\includegraphics[width=8cm]{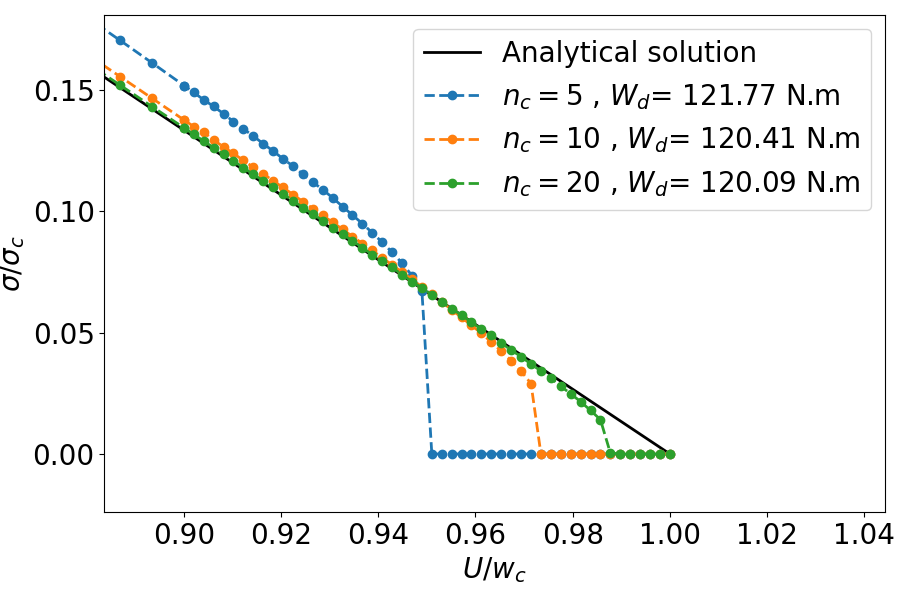}
\end{tabular}
\end{center}
\caption{Phase-field: Zoom around $U = \wc$ during X-Mesh simulations for different values of $\nc$. The stress discontinuously drops to zero for $U^{*} < U$.}
\label{fig:XMeshSnapBack}
\end{figure}

To check the convergence of the results, the stress versus imposed displacement curves are plotted for different values of $n_c$ in Figure \ref{fig:PhaseFieldXMeshAndFixedMeshConvergenceSigU}.  Even with a very fine fixed mesh, the bar never breaks, whereas the results with X-Mesh are almost independent of the mesh element size.
We also made a ``zoom'' around $U = \wc$ by adding more increments of $U$ in Figure \ref{fig:XMeshSnapBack}. This provides a clearer view of the stress drop phenomenon observed in Figure \ref{fig:phaseField1DDispAndDamage} (a). These figures show that as $U$ increases but remains far from $\wc$, $\sigma$ gradually decreases, following the analytical solution curve. However, at a certain value of $U$, $\sigma$ suddenly drops to zero (corresponding to $d_0$ suddenly ``jumping'' from a value strictly less than 1 to 1). Using smaller increments of $U$ confirms that this drop occurs abruptly and discontinuously.
The evolution of $(h,u,d)$ is not continuous with respect to $U$ for X-Mesh.
For some values of $U^{*}$, a jump occurs. As the mesh is refined, $U^{*}$ is closer and closer to $\wc$ and the jump gets closer and closer to zero. A possible explanation for this phenomenon will be provided in a later section.

\begin{figure}
\begin{center}
\includegraphics[width=9cm]{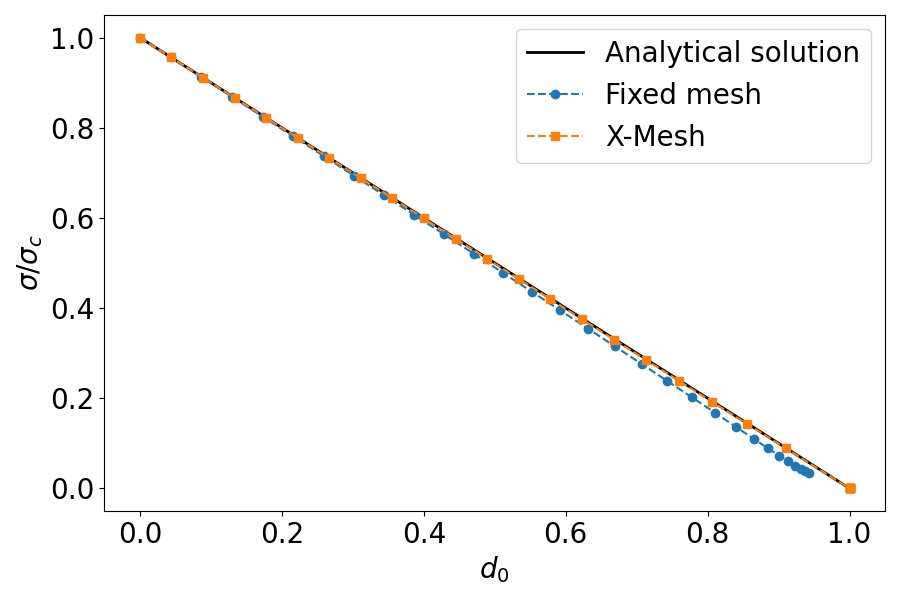}
\end{center}
\caption{Phase-field: Stress $\sigma$ as a function of $d_0$.}
\label{fig:PhaseFieldXMeshAndFixedMeshComparisonSig}
\end{figure}

Finally, figure \ref{fig:PhaseFieldXMeshAndFixedMeshComparisonSig} shows the stress $\sigma/\sigc$ , computed using equation \eqref{eq:sigDisc} as a function of $1-d_0$. As can be seen, for X-Mesh this curve is a linear curve, confirming a property of the numerical solution that was established in section \ref{sec:xmesh}, namely that the stress satisfies the analytical relation \eqref{eq:exactPFStress} when the mesh is optimised, which is not the case with a fixed mesh.

\subsection{Lip-field analysis}

\begin{figure}
\begin{center}
\includegraphics[height=6cm]{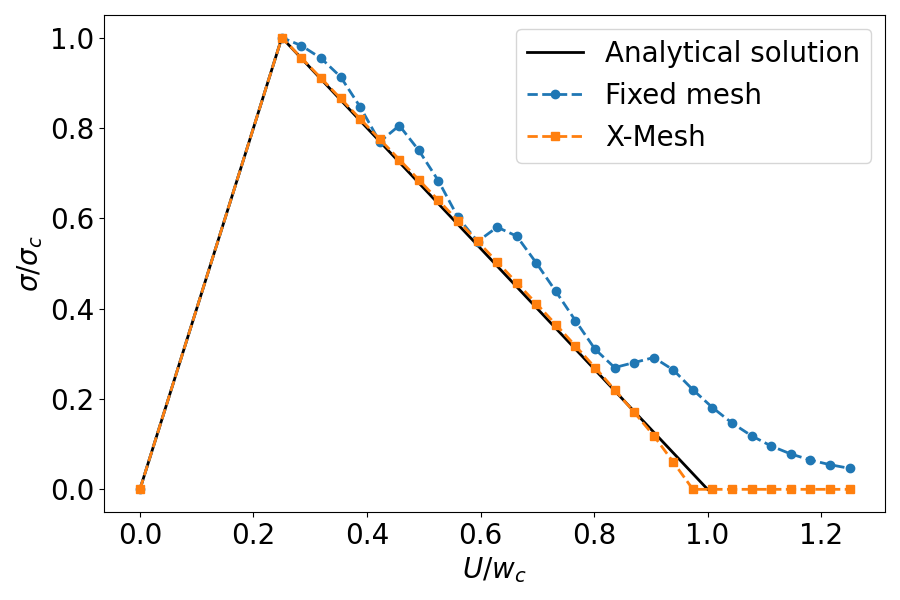}
\\ (a) Stress versus imposed displacement (lip-field)
\\
\begin{tabular}{cc}
\includegraphics[width=8cm]{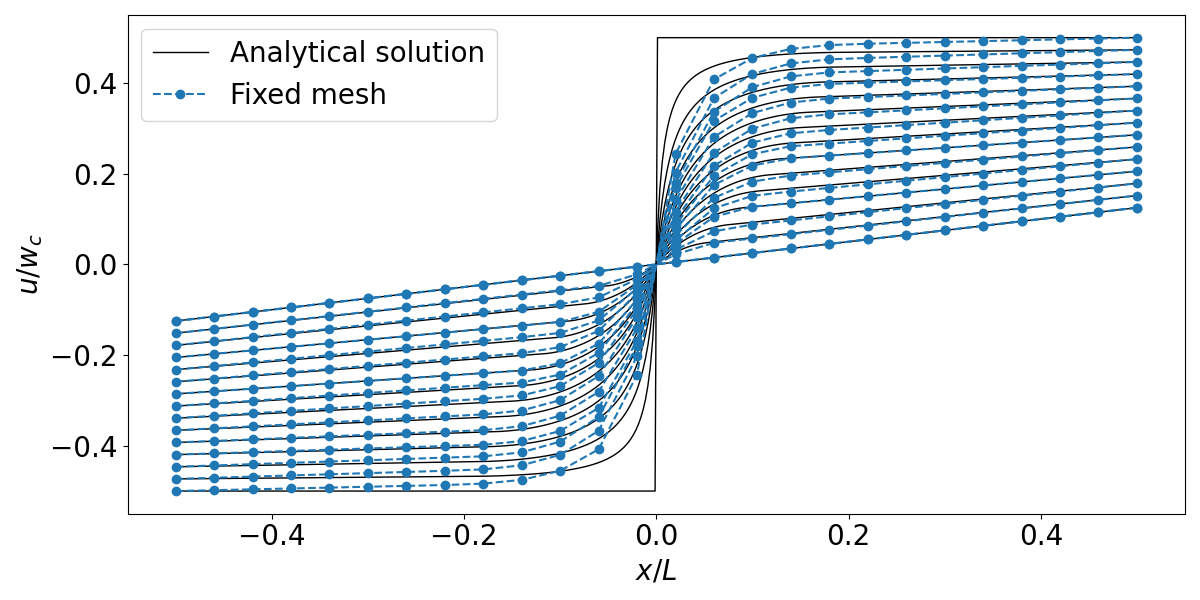}
& \includegraphics[width=8cm]{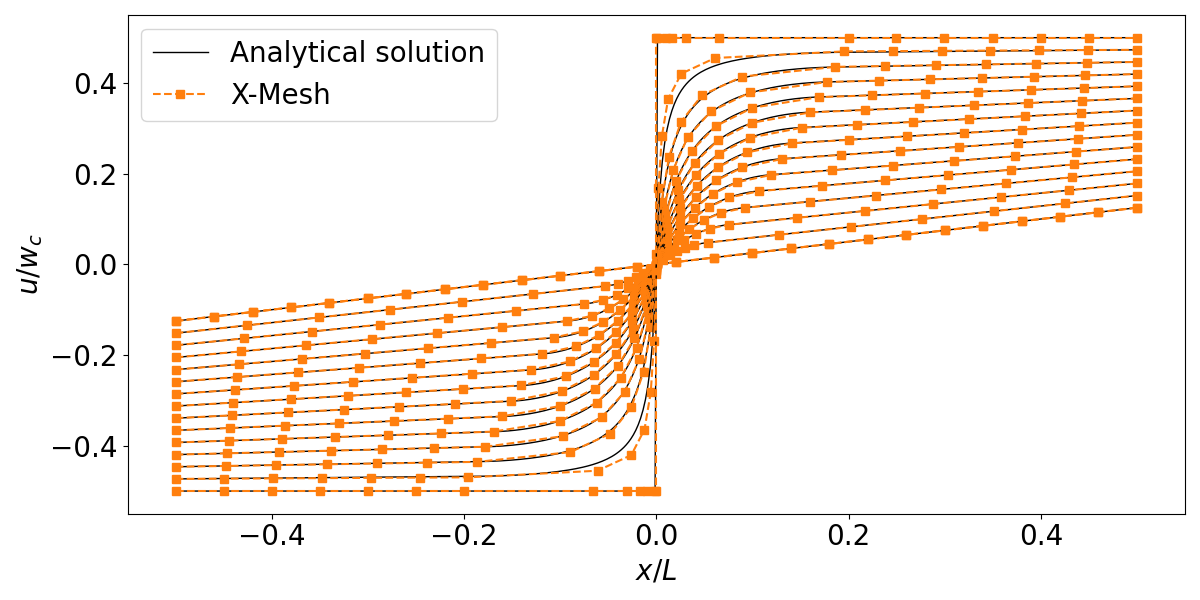}
\\
\end{tabular}
\\ (b) Displacement fields (lip-field)
\\
\begin{tabular}{cc}
\includegraphics[width=8cm]{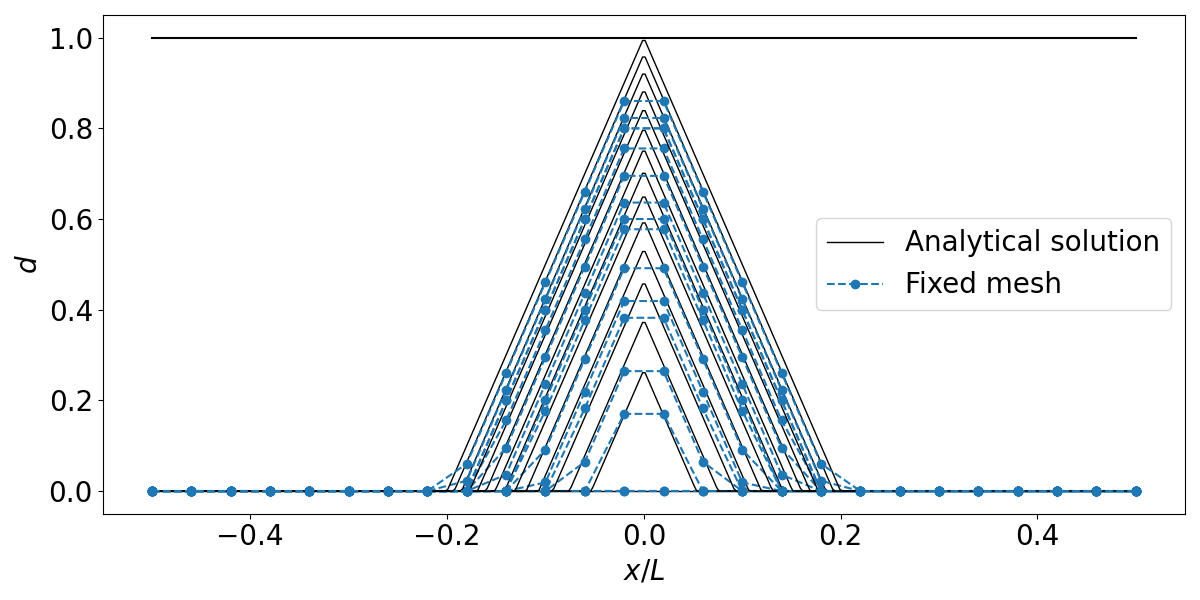}
& \includegraphics[width=8cm]{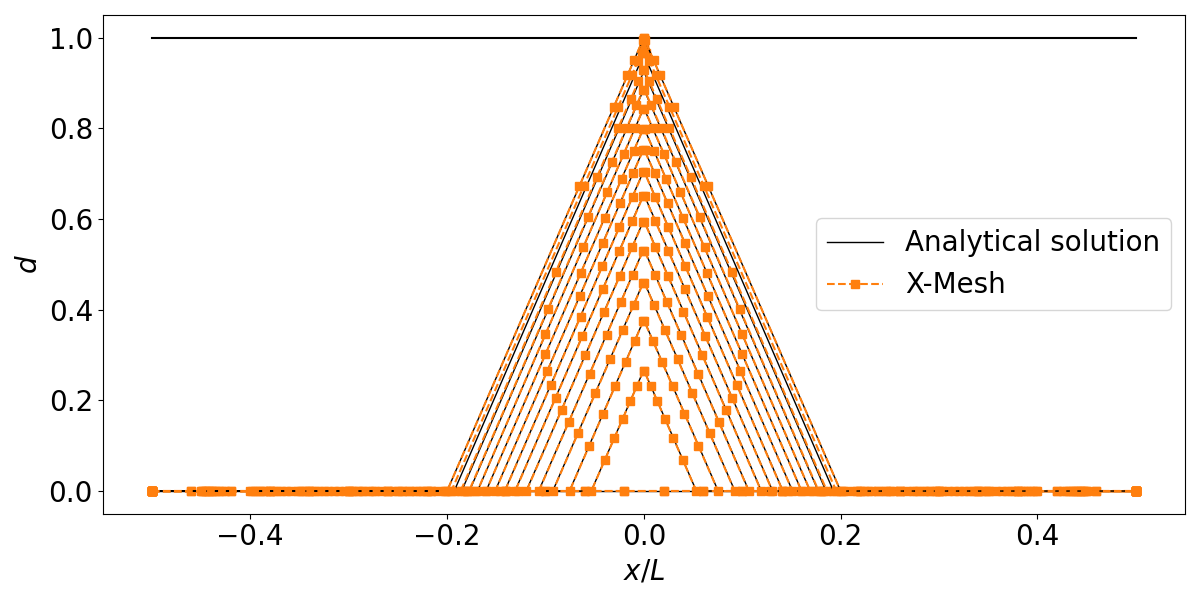}
\\
\end{tabular}
(c) Damage fields (lip-field)
\begin{tabular}{cc}
\includegraphics[height=5cm]{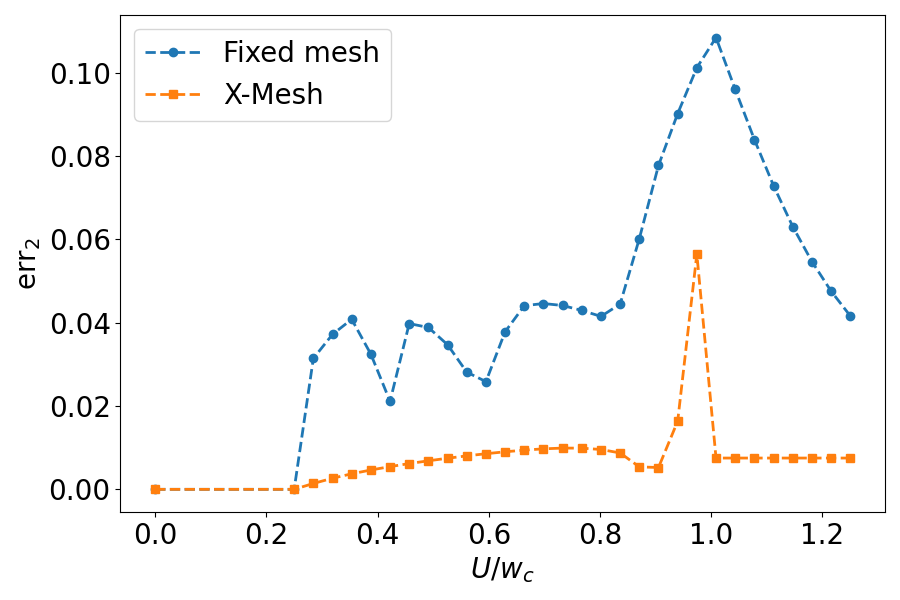}
& \includegraphics[height=5cm]{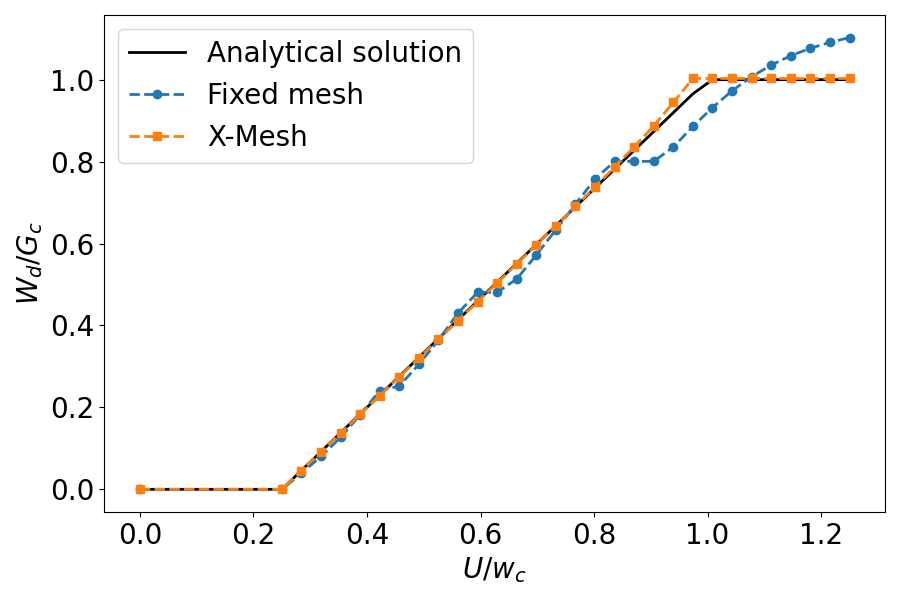}
\\
(d) $L_2$ error on displacement (lip-field) & (e) Dissipated energy (lip-field)
\end{tabular}
\end{center}
\caption{Lip-field: results obtained with a fixed mesh and X-Mesh, for five elements per $\lc$.}
\label{fig:lipField1DDispAndDamage}
\end{figure}

\begin{figure}
\begin{center}
\includegraphics[width=12cm]{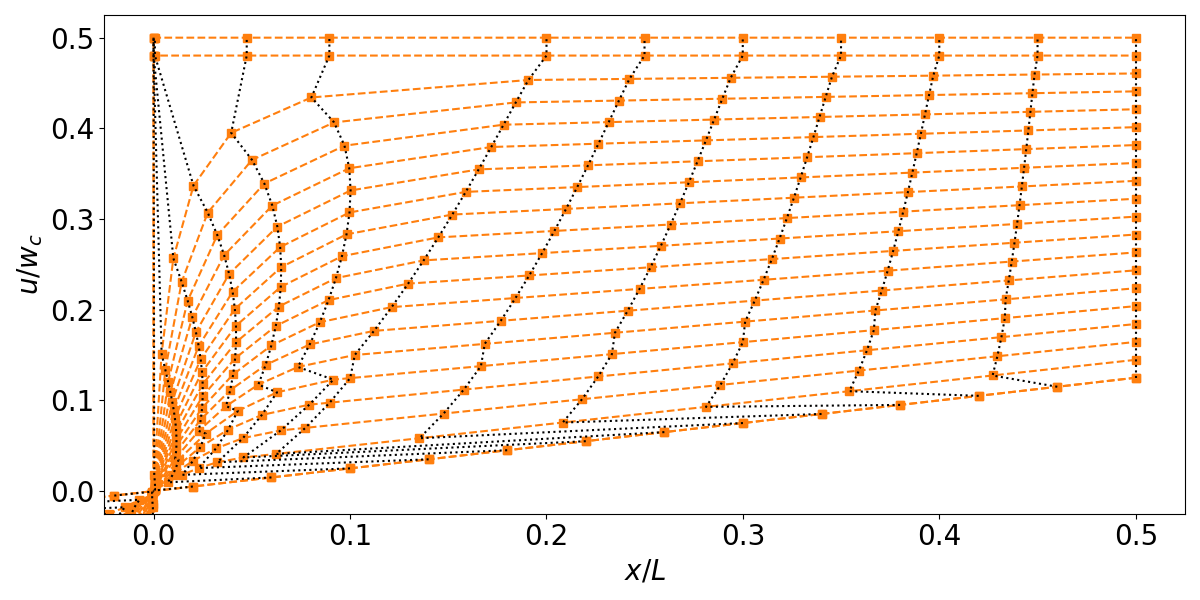}
\end{center}
\caption{Lip-field: Nodes displacement during X-Mesh computation.}
\label{fig:LipFieldXMeshNodesTrajectories}
\end{figure}

This section presents the results obtained using the lip-field approach. The stress versus imposed displacement curves and the displacement and damage fields, are plotted in Figure \ref{fig:lipField1DDispAndDamage}. Figure \ref{fig:LipFieldXMeshNodesTrajectories} shows the nodes trajectories during the X-Mesh simulation. These results are similar to those obtained using phase-field: the bar breaks with X-Mesh, but not with a fixed mesh. The size of the central elements goes to zero as $d_0$ goes to 1. The X-Mesh results in Figure \ref{fig:lipField1DDispAndDamage} (a) shows that as with phase-field, the evolution of $(h,u,d)$ is also discontinuous. Some differences can also be observed compared to phase-field. For the fixed mesh results, the curve in Figure \ref{fig:lipField1DDispAndDamage} (a) shows some kind of oscillations. In fact, closer inspection of these oscillations reveals that the ascending phases correspond to elastic reloadings where damage does not evolve. This is confirmed by Figure \ref{fig:lipField1DDispAndDamage} (e) which shows that dissipation does not evolve during certain time steps. We will explain this phenomenon in what follows. Note that the X-mesh curve does not exhibit this issue. Also, unlike phase-field where the damage field has a zero slope at $x = 0$, except when $d_0 = 1$ when it has a kink point, the damage field with lip-field has a kink throughout the entire breaking process. (The X-Mesh recovers the features of the analytical solution mentioned in section \ref{sec:model}). Therefore, the mesh nodes tend to create a central element of very small size from the beginning of the simulation, reaching a zero size when $d_0 = 1$ (Figures \ref{fig:lipField1DDispAndDamage} (b) and (c)). With phase-field, $h_0$ decreases to zero more progressively.

Finally, with a fixed mesh, the damage slope is $\pm 1/\lc $ on all elements of the damaged zone (elements for which $\dmid > 0$), except on the outermost elements of the damage zone. We will refer to these elements as $\nc$ in what follows. They are the only elements such that $\dmid_{\nc} > 0$, but $d$ is zero on one of their two nodes. Conversely, with X-Mesh, one node of $\nc$ moves such that the slope of $d$ is exactly $\pm 1/\lc $, which confirms the analysis made in section \ref{sec:xmesh}.

\begin{figure}
\begin{center}
\begin{tabular}{cc}
\includegraphics[width=8cm]{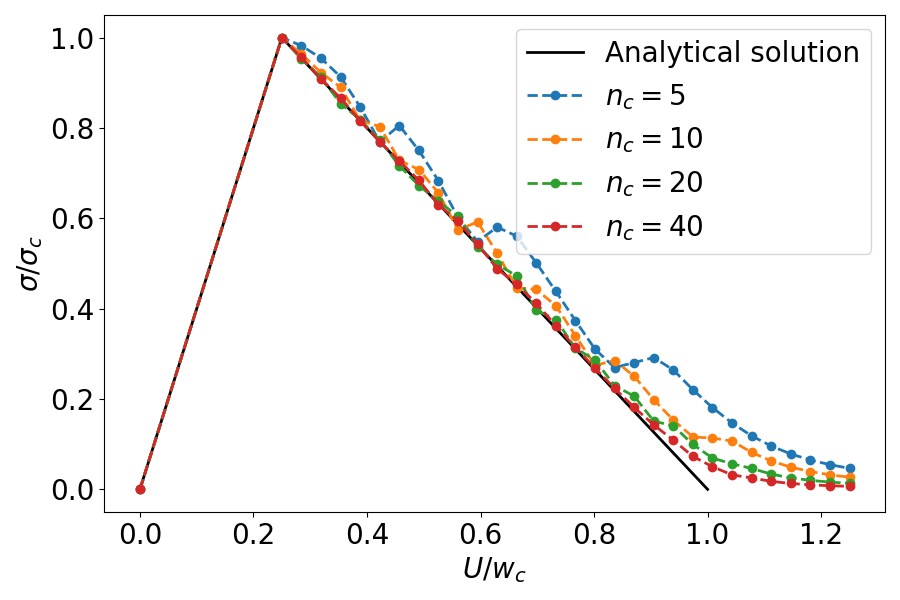}
&
\includegraphics[width=8cm]{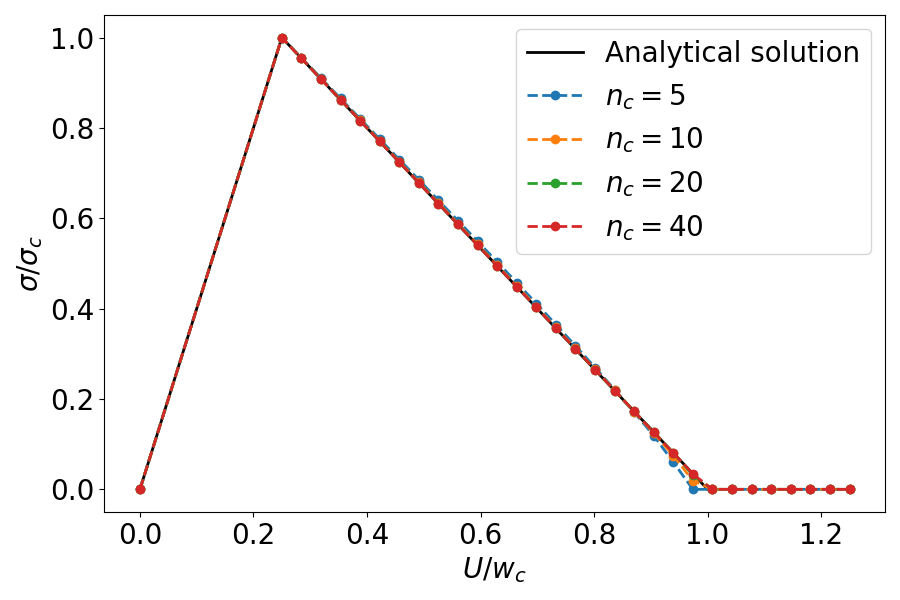}
\end{tabular}
\end{center}
\caption{Lip-field: Stress versus imposed displacement curves obtained with a fixed mesh (left) and for X-Mesh (right) for different values of $n_c$.}
\label{fig:LipFieldXMeshAndFixedMeshConvergenceSigU}
\end{figure}

\begin{figure}
\begin{center}
\includegraphics[width=9cm]{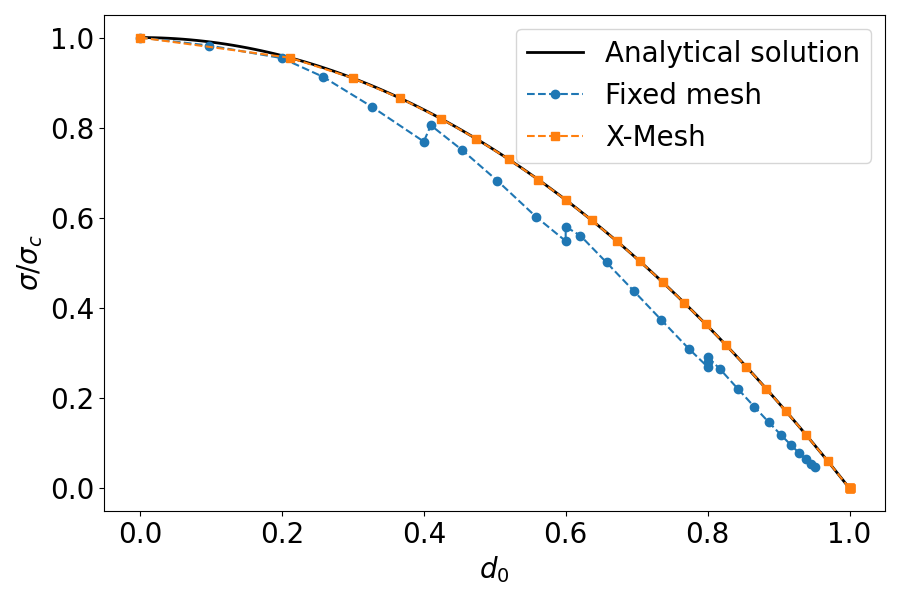}
\end{center}
\caption{Lip-field: Stress $\sigma$ as a function of $d_0$.}
\label{fig:LipFieldXMeshAndFixedMeshComparisonSig}
\end{figure}

To check convergence, the stress versus imposed displacement curves are plotted for different values of $n_c$ in Figure \ref{fig:LipFieldXMeshAndFixedMeshConvergenceSigU}. As with phase-field, the results obtained with lip-field converge to the analytical curve as the mesh is refined. In particular, with a fixed mesh, the amplitude of the elastic reloading decreases as the size of the mesh elements decreases.
Also, Figure \ref{fig:LipFieldXMeshAndFixedMeshComparisonSig} shows that with X-Mesh, the stress satisfies the analytical equation \eqref{eq:exactLipStress}.

\begin{figure}
\begin{center}
\begin{tabular}{cc}
\includegraphics[width=8cm]{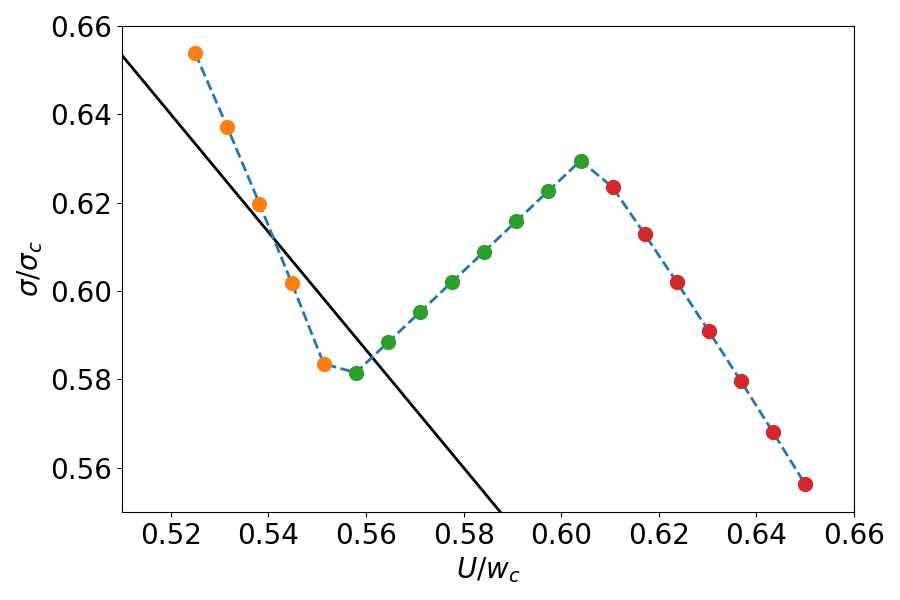}
&
\includegraphics[width=8cm]{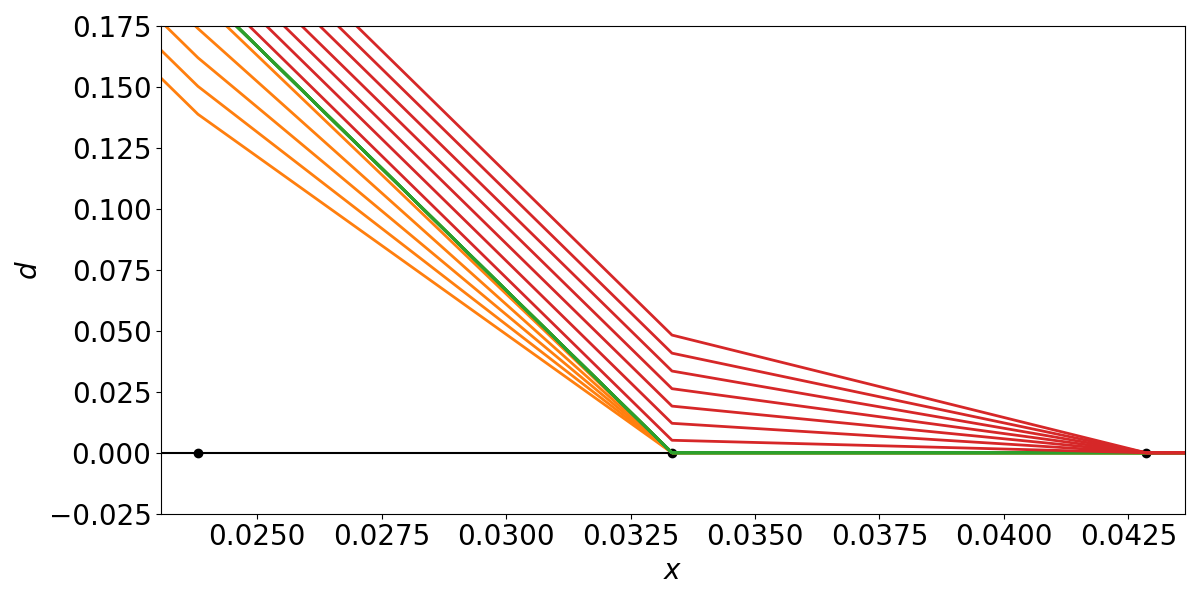}
\\
(a) & (b)
\\
\includegraphics[width=8cm]{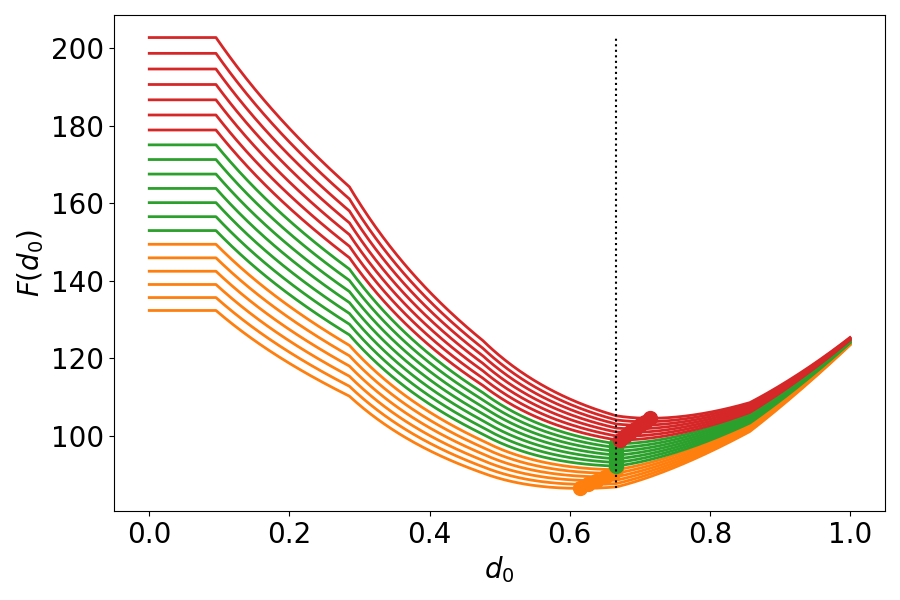}
&
\includegraphics[width=8cm]{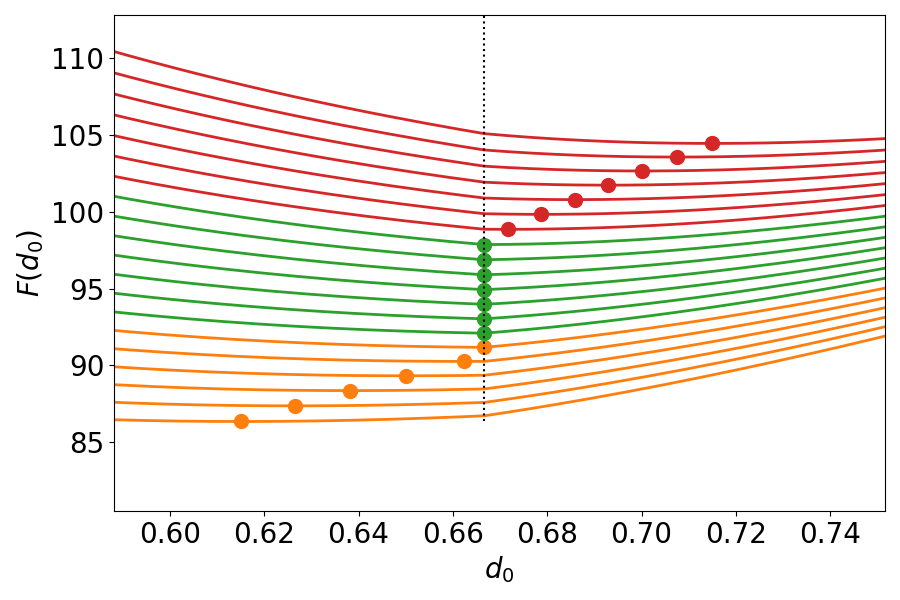}
\\
(c) & (d)
\end{tabular}
\end{center}
\caption{Lip-field: Elastic reloading phenomenon observed with lip-field on fixed meshes. (a) Zoom on stress vs imposed displacement curve around elastic reloading. (b) Damage field around the last element where $d > 0$. (c) Plots of $F(d_0)$. (d) Plots of $F(d_0)$, zoomed in around the discontinuity.}
\label{fig:LipFieldDebug}
\end{figure}

To understand the phenomenon of elastic reloading observed in Figure \ref{fig:lipField1DDispAndDamage} (a), we focus on the behaviour of the solution around the one of these elastic reloading in Figure \ref{fig:LipFieldDebug}. Each coloured dot in Figure \ref{fig:LipFieldDebug} (a) corresponds to an increment of $U$, which has a damage field counterpart of the same colour on \ref{fig:LipFieldDebug} (b) (which is a zoom around $\nc$). It seems that the reloading (green dots in Figure \ref{fig:LipFieldDebug} (a)) occurs when damage becomes non-zero on a new finite element. As expected, this corresponds to an increment where damage does not evolve (all the green curves are superimposed in Figure \ref{fig:LipFieldDebug} (b)). Also, during the elastic reloading, it seems that the damage slope on $\nc$ is exactly $\pm \lc$. To gain a deeper understanding of this phenomenon, we can observe that all the damage fields plotted in Figure \ref{fig:lipField1DDispAndDamage} (b) can be expressed as

\begin{equation}
 \label{eq:lipNumericalDamage}
d_i = \left< d_0 - \frac{x_i}{\lc} \right>_+
\end{equation}
where $\left<   \blacksquare  \right>_+$ is the positive part of ``$\blacksquare$''. By substituting \eqref{eq:lipNumericalDamage} into \eqref{eq:F}, we can express the objective function $F$ as a function of one scalar variable $d_0$ and plot it for a given value of $U$, as shown in Figures \ref{fig:LipFieldDebug} (c) and (d), which is a zoom around the particular value of $d_0$ that corresponds to the elastic reloading in this particular example. The function $F(d_0)$ is continuous with respect to $d_0$, but its derivative is not. It can be seen that $F(d_0)$ has the shape of a ``basin'', but with some corners. As the value of $U$ increases, the ``basin'' rotates. For certain values of $U$ (the orange and red dots in Figures \ref{fig:LipFieldDebug} (c) and (d)), the minimum is on a smooth part of the basin, so its moves continuously inside its bottom. However, for certain values of $U$, the minimum becomes ``stuck'' in a corner of the basin, causing $d_0$ to stagnate despite $U$ increasing.

\subsection{Lip-field five elements analysis}

\begin{figure}
\begin{center}
 \includegraphics[width=9cm]{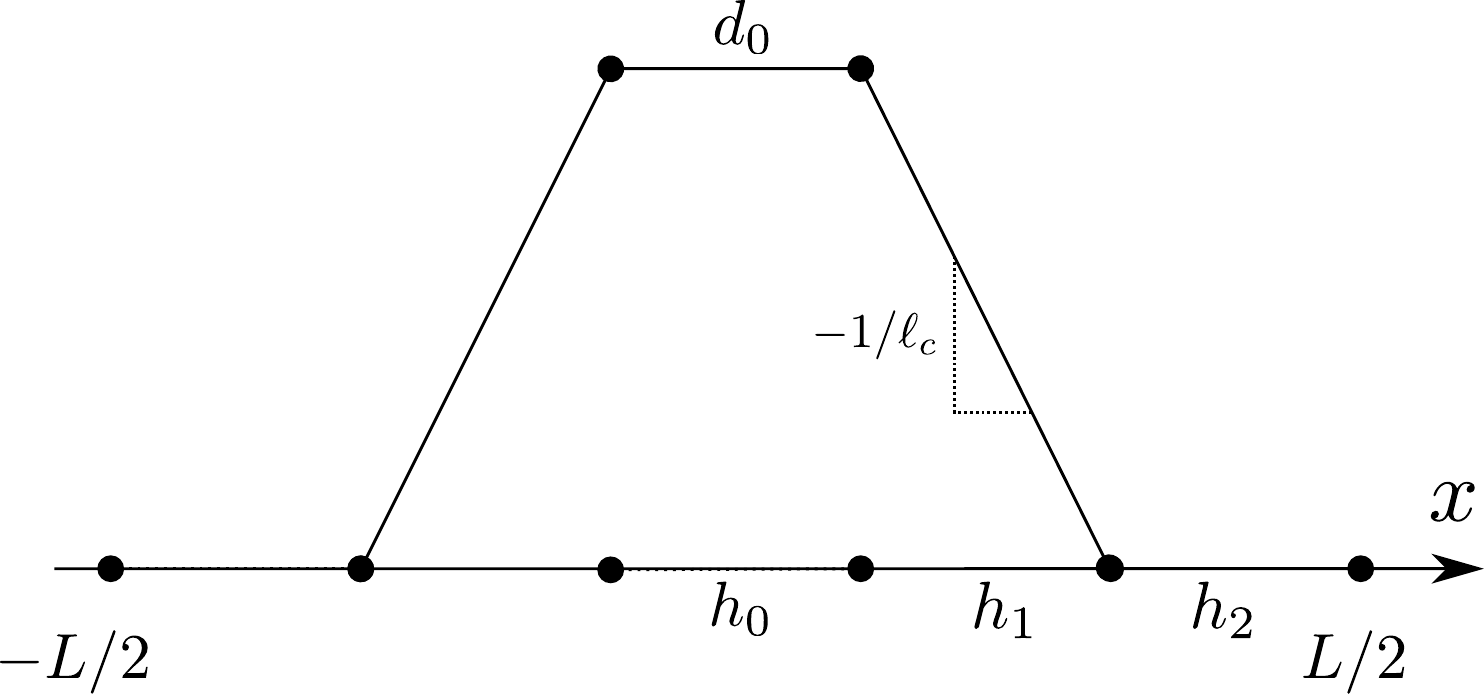}
\end{center}
\caption{Lip-field five elements case.}
\label{fig:lip1DFiveElements}
\end{figure}
To understand the ``jump'' observed in the previous section when $d_0$ tends to 1, we will consider a simple lip-field X-Mesh example with only five elements, as illustrated in Figure \ref{fig:lip1DFiveElements}. We assume that damage is non-zero only at the nodes of the central element. According to the analysis made in section \ref{sec:xmesh}, the damage slope must be exactly $\pm 1/\lc$ when the mesh is optimised, which establishes a relationship between $d_0$ and $h_1$. Additionally, the sum of the lengths of the mesh elements must equal the length of the bar, yielding the following relations:
\begin{equation}
 h_1 = d_0 \lc, \quad h_2 = L/2 - (h_0/2 +d_0 \lc).
\end{equation}
The following expression for the function $F$ can be obtained for a five-element mesh, denoted by $\Ffive$, which depends only on $d_0$ and $h_0$:
\begin{align}
\label{eq:FFiveElem}
    \Ffive(d_0,h_0) & = F_e(U) \Kfive(d_0,h_0) + \frac{\Gc}{\lc} (h_0 \alphaL(d_0) + 2 d_0 \lc \alphaL(d_0/2)) \\
    \Kfive(d_0,h_0) & = L \left(  h_0 \omegaL^{-1}(d_0) + 2 d_0 \lc \omegaL^{-1}(d_0/2) + L - (h_0 + 2 d_0 \lc) \right)^{-1}.
\end{align}
For the purposes of what follows, we will assume that $\gamma = \frac{1}{2}$. We can then express $\Ffive$ as a function of $d_0$ only. To achieve this, we first use the expression given in equation \eqref{eq:exactLipStress} and the definition of $\sigma$:
\begin{equation}
 E U = \sigc (1-d_0^2)L \Kfive^{-1}
\end{equation}
which gives (after some calculations) the expression of $h_0 d_0$:
\begin{equation}
\label{eq:hZerodZeroFiveElem}
 h_0 d_0 =\frac{E U (1-d_0^2)}{4 \sigc} - \frac{L}{4} (1-d_0^2)^2 - \frac{ \lc d_0^2 (1-d_0^2)^2}{(1-(d_0/2)^2)^2}
\end{equation}
while the expression of $F$ simplifies to
\begin{equation}
\label{eq:FhZerodZeroFiveElem}
 \Ffive(d_0,h_0 d_0) = \frac{1}{2} U \sigc (1-d_0^2) + 2 \frac{\sigc^2}{E} ( h_0 d_0 + d_0^2 \lc).
\end{equation}
Combining the two equations above gives us the result that $\Ffive$ is a function of $d_0$ only.
Assuming a linear evolution of $d$, we can determine the expression of $\Finf$, which corresponds to the expression of $F$ that would be obtained by discretising the bar with an infinite number of elements. This is obtained by analytically integrating the different terms of $F$ in \eqref{eq:lipFieldFunctional}:
\begin{equation}
 \Finf(d_0) = F_e(U) \frac{L \omega(d_0)}{h_0 + 2 d_0 \lc \frac{\omega(d_0)}{\omega(d_0/2)} + \omega(d_0)(L-(h_0+2 d_0 \lc)) }
            + \Gc d_0^2.
\end{equation}

\begin{table}
\centering
\begin{tabular}{llcc}
  \hline
  Properties  & Units     &  Symbol    &  Value   \\ \hline
  Bar length & m & $L$ & 0.22 \\
  Regularization length & m & $\lc$ & 0.2 \\
  Young's modulus & Pa &  $E$    &  $ 3 \cdot 10^{10}$    \\
  Fracture toughness & N/m & $G_c$ & $120$ \\
  Critical tensile stress & Pa & $\sigc$ & $3 \cdot 10^6$ \\
  \hline
 \end{tabular}
 \caption{Dimensions an material properties for lip-field five elements example.}
 \label{table:lipFiveElementsMaterialProperties}
\end{table}

\begin{figure}
 \centering
 \begin{tabular}{m{0.5cm}m{11cm}}
  (a) &
 \includegraphics[width=11cm]{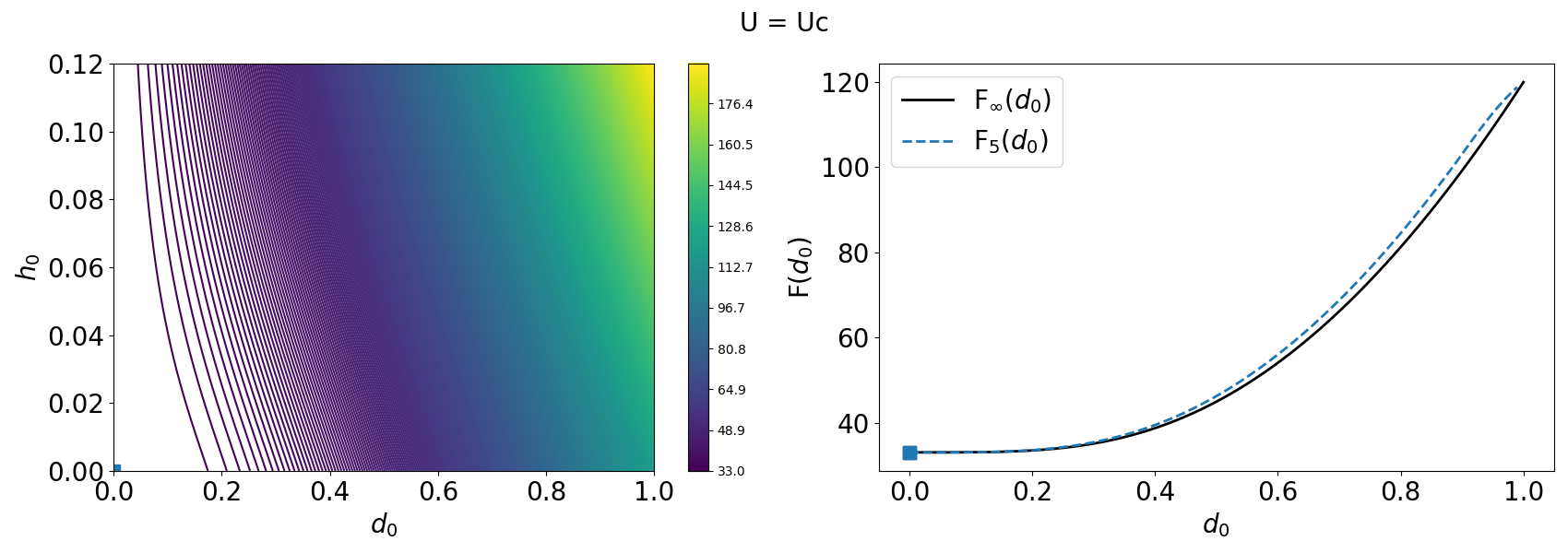} \\
  (b) &
 \includegraphics[width=11cm]{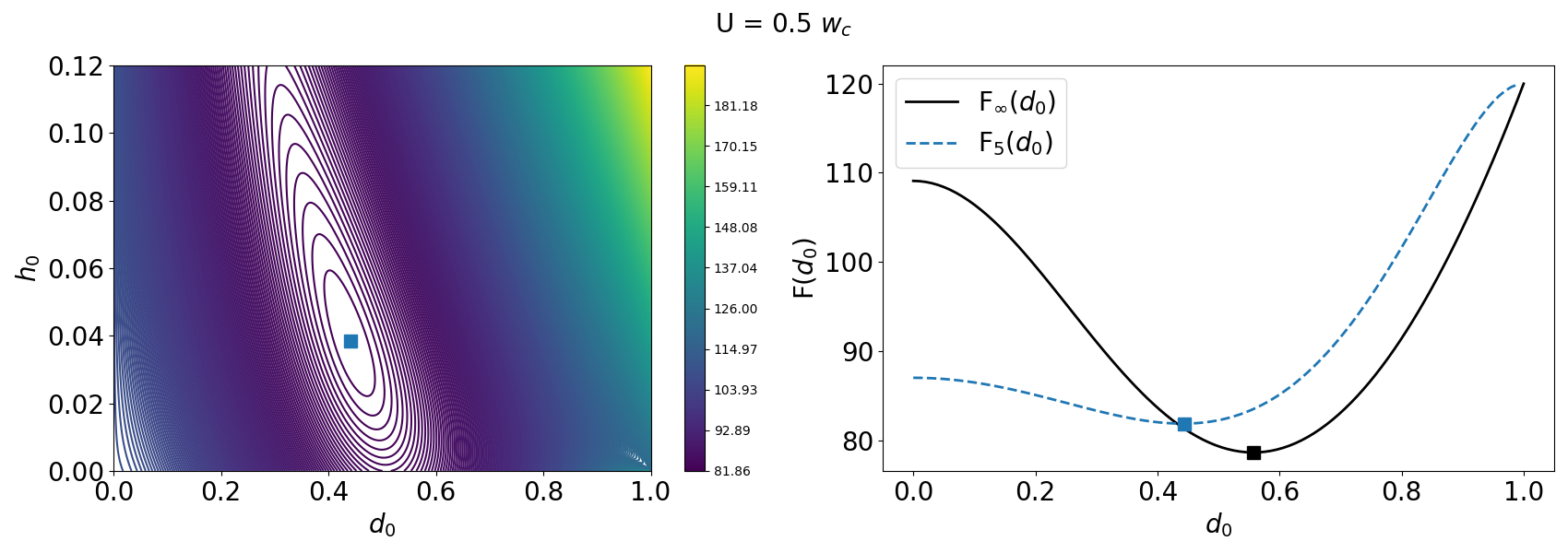} \\
  (c) &
 \includegraphics[width=11cm]{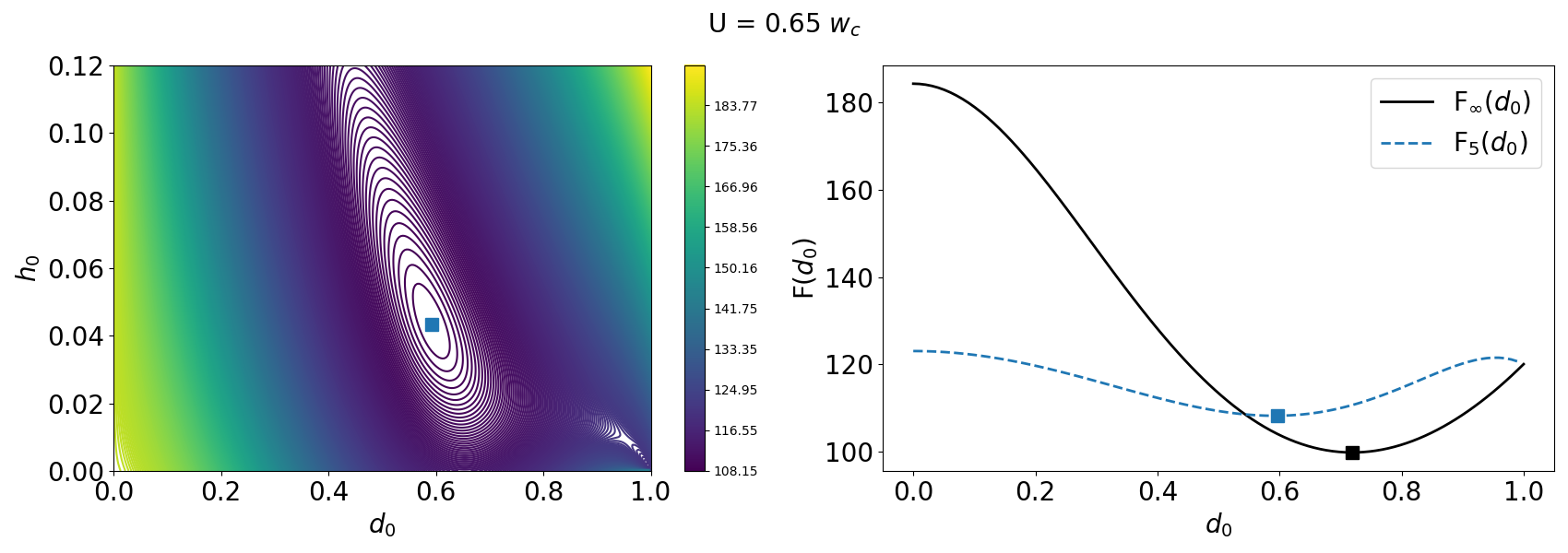} \\
  (d) &
 \includegraphics[width=11cm]{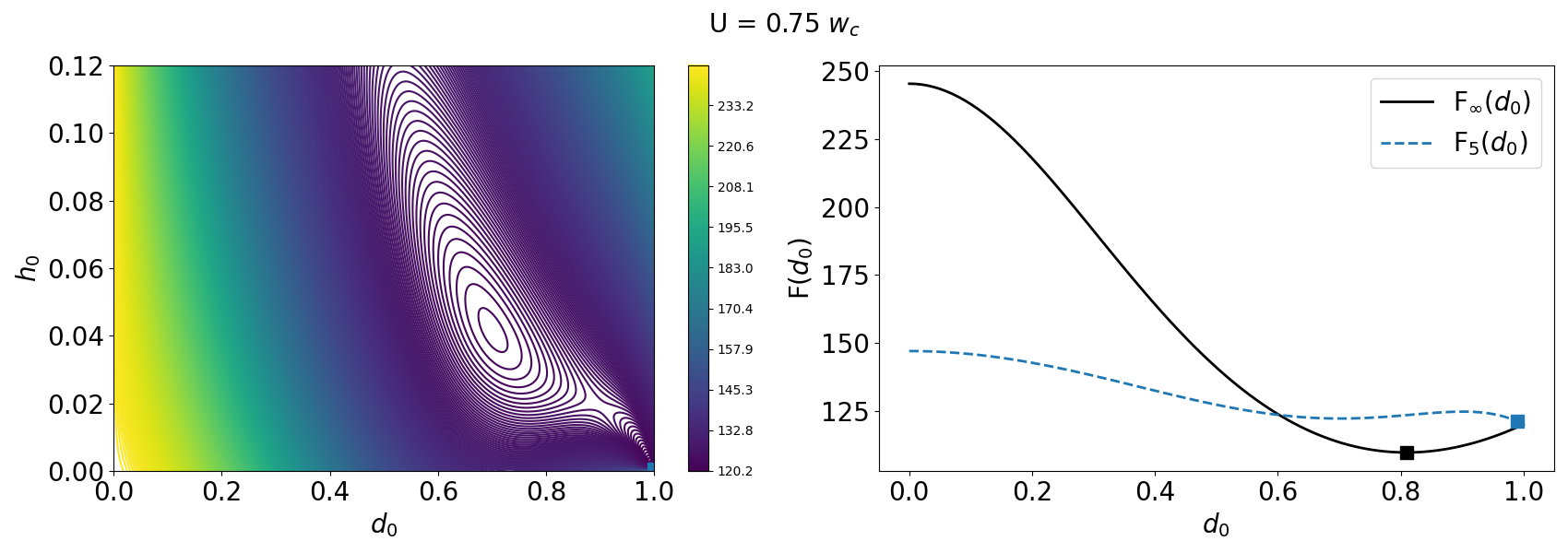} \\
  (e) &
 \includegraphics[width=11cm]{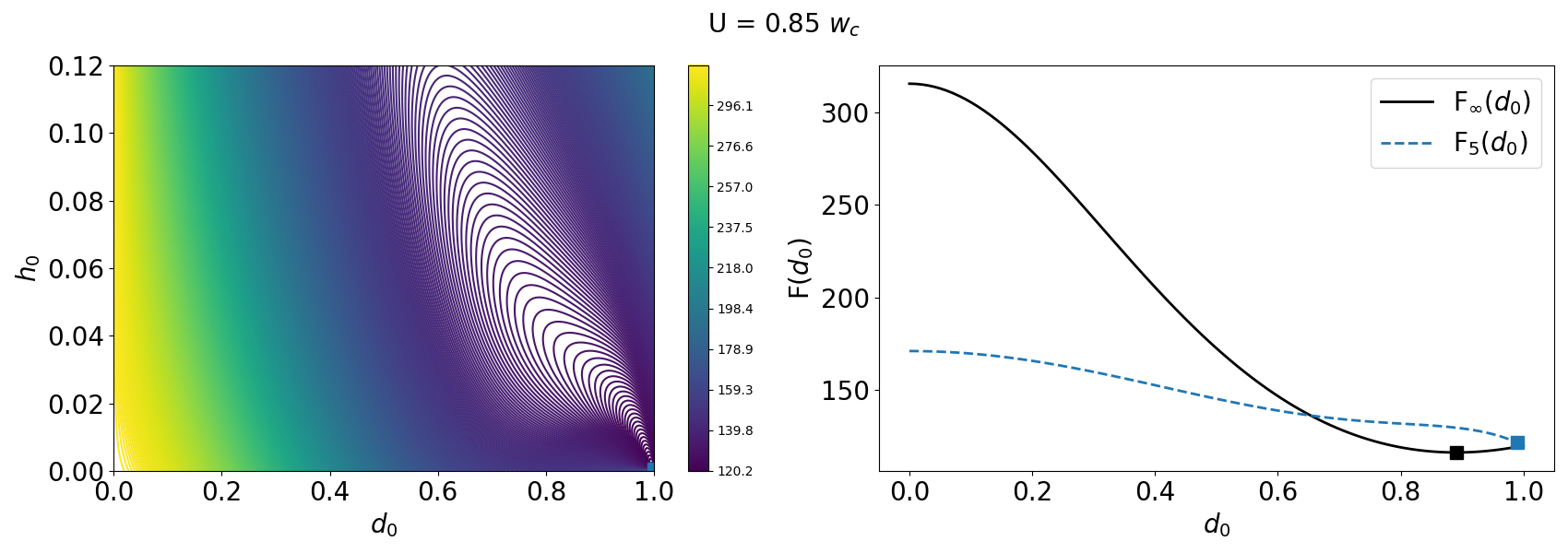} \\
 \end{tabular}
  \caption{Results for the lip-field five elements example. Left: $\Ffive$ as a function of $d_0$ and $h_0$ (equation \eqref{eq:FFiveElem}. Right: $\Ffive$ as a function of $d_0$ only (obtained using equations \eqref{eq:hZerodZeroFiveElem} and \eqref{eq:FhZerodZeroFiveElem} and $\Finf$. The blue square corresponds to the global minimum of $\Ffive$, and the black square to the global minimum of $\Finf$.}
 \label{fig:lipFiveElementsResults}
\end{figure}
We plot $\Ffive(d_0,h_0)$, $\Ffive(d_0)$ and $ \Finf(d_0)$ for the values of the numerical parameters given in table \ref{table:lipFiveElementsMaterialProperties} in Figure \ref{fig:lipFiveElementsResults}, at five different stages as the imposed displacement $U$ increases:
\begin{enumerate}[label=(\alph*)]
 \item \label{stage:lipFiveElemA} The first stage $U < U_c$, corresponds to no damage ($d_0 = 0$). Interestingly, Figure \ref{fig:lipFiveElementsResults} (a) shows that the corresponding $h_0$ is 0, which is sufficient to represent the behaviour of an undamaged, elastic bar.
 \item \label{stage:lipFiveElemB} A second stage where $\Ffive$ has only one global minimum $d_{0, \text{min}} < 1$. As $U$ increases, the value of $d_{0, \text{min}}$ increases gradually and continuously.
 \item \label{stage:lipFiveElemC} A third one where $\Ffive$ has two local minima: one for $d_{0, \text{min}} < 1$ and one for $d_0 = 1$. At this stage $d_{0, \text{min}}$ remains the global minimum.
 \item \label{stage:lipFiveElemD}  A fourth stage where $\Ffive$ still has two local minima $d_{0, \text{min}} < 1$ and 1, but now the global minimum is now at $d_0 = 1$.
 \item \label{stage:lipFiveElemE}  A fifth stage where $\Ffive$ only has one minimum for $d_0 = 1$.
\end{enumerate}
The resolution during stage \ref{stage:lipFiveElemB} will not cause any particular problems, since $\Ffive$ only has one global minimum. For stage \ref{stage:lipFiveElemC}, it should be possible to achieve a continuous increase in $d_{0, \text{min}}$, if the solver used always finds the global minimum. However, with stage \ref{stage:lipFiveElemD}, $d_0 = 1$ suddenly becomes the global minimum of $\Ffive$. This problem can be avoided by searching for a minimum in the neighbourhood of the previous solution, thus ensuring continuity in the evolution of $d_{0, \text{min}}$ between stages \ref{stage:lipFiveElemC} and \ref{stage:lipFiveElemD}. The problematic stage is stage \ref{stage:lipFiveElemE}: indeed, the local minimum $d_{0, \text{min}}$ ``disappears''. The solution then jumps from $d_{0, \text{min}} < 1$ to $d_0 = 1$. Note that this occurs for a value of $U$ which is still far from $\wc$. Not only does $d_0$ suddenly goes to 1, but also $h_0$ suddenly goes to 0 as can be seen in Figures \ref{fig:lipFiveElementsResults} (d) and (e), left. This explains why the stress suddenly drops to zero in the results of the previous sections, assuming that this reasoning is also valid for phase-field, and for any number of finite elements in the mesh. Note that $\Finf$ do not have this issue: whatever the value of $U$, there is only one global minimum, which increases continuously as $U$ increases.

\section{Discussion and conclusions}
\label{sec:conclusion}

In this paper, we focus on a one-dimensional fracture problem modelled using phase-field and lip-field approaches. Both approaches consist of finding the minimum of an incremental potential function with respect to displacement and damage fields. The difference lies in how the regularisation length $\lc$, which is necessary to avoid spurious mesh dependency, is introduced: through a gradient term in the expression of the incremental functional for phase-field and by a Lipschitz restriction on the damage variable for lip-field. This work examined using the X-Mesh approach to enhance the outcomes achieved with phase-field and lip-field. This approach involves moving the nodes of the finite element mesh; the movement of the nodes is determined by considering the nodal coordinates as an unknown in the optimisation problem. From this one-dimensional study, we were able to draw the following conclusions:
\begin{itemize}
 \item Using a fixed mesh, the bar does not break for both phase-field and lip-field (i.e. the damage on the central element $d_0$ never goes to 1). Conversely, with X-Mesh, $d_0$ goes to one, while the size of the central element $h_0$ goes to zero.
 \item It is not only the nodes of the central element that move, but also all the elements of the mesh. This improves the quality of the solution globally, not only when $d_0 \rightarrow 1$. In particular, with the lip field, it is possible to achieve a slope in the damage field of exactly $\pm 1/\lc$, eliminating the elastic reloadings that can be observed with a fixed mesh.
 \item A relatively low number of elements achieves very good accuracy. In particular, the stress versus imposed displacement relationship is almost independent of the mesh element size. Using a moving mesh also allows the exact stress versus $d_0$ relation to be recovered for both phase-field and lip-field.
 \item However, with X-Mesh, the evolution of $d_0$ with respect to the imposed displacement $U$ is not continuous, dropping suddenly to one for a value of $U$ smaller than the critical value. This drop is due to the non-convexity of the incremental potential to optimise, which leads to the appearance/disappearance of local minima. It should be noted that this drop occurs at a value of $U$ closer to the critical value as the mesh is refined, indicating some degree of convergence towards the continuous model, for which the evolution of $d_0$ is continuous.
\end{itemize}
Finally, the main contribution of X-Mesh to fracture modelling is to adjust the finite element mesh nodes so that the material can be fully broken and displacement jumps can be modelled, which is not possible with a fixed mesh. This involves elements of size zero, a phenomenon that arises naturally from optimising the incremental potential with respect to mesh element sizes. Future work will focus on extending to 2D/3D; the main challenge here is to avoid negative Jacobian elements, which is trivial in 1D but more complex in 2D/3D. Although X-Mesh allows for better simulation with a given number of nodes than a fixed mesh, this comes at the cost of a much slower solver. Optimising node locations is rather slow (X-Mesh computations require between two and ten times as many iterations as fixed mesh computations), and we would like to draw attention to a similar observation by Maugin in their state-of-the-art paper  \cite{maugin2013}: “One cannot avoid mentioning here problems posed by the rather slow convergence of the implied iterative procedure”. However, once a suitable solver has been found, we expect the nodes to move naturally to the location of the crack, forming layers of zero-size elements. Additionally, the stress drop phenomenon observed in 1D must be carefully studied.

\section*{Declaration of competing interest}

The authors do not work for, advise, own shares in, or receive funds from any organisation that could benefit from this article, and have declared no affiliation other than their research organisations.

\section*{Acknowledgements}

This project has received funding from the European Research Council (ERC) under the European Union’s Horizon research and innovation programme (Grant agreement No. 101 071 255).

\appendix

\section*{Appendix}
\label{appendix}

In this section we present the analytical solutions of the phase-field and lip-field models described in section \ref{sec:model}.

\section{Phase-field}
\label{appendix:phasefield}

Regarding the damage profile over the bar, it is given implicitly by \cite{wu2017}
\begin{equation}
 \frac{\dint d}{\dint x} = -\frac{1}{\lc}  H(d,d_0), \quad
  H(d,d_0)
 = \sqrt{ (2d - d^2) \left( 1 - \left(\frac{1-d_0}{1-d} \right)^2 \right) } \label{eq:gradex}.
\end{equation}
The profile $\dP(x)$ is obtained by inverting the above equation. An explicit solution exists for $d_0=1$:
\begin{equation}
 \dP(x) = 1-\sin\left(\frac{x}{\lc}\right).
\end{equation}
Also, it turns out that the width of the damaged zone obtained as:
\begin{equation}
 \ell = \lc \int_0^{d_0} H^{-1}(\tilde{d},d_0) \dint \tilde{d}
\end{equation}
does not depend on $d_0$:
\begin{equation}
\label{eq:PFDamagedBandWidth}
 \ell = \frac{\pi}{2} \lc.
\end{equation}
This means that, as soon as the damage starts, it will occupy its full extent, which will not vary.
We note that the gradient of $d$ at $x=0$ is always 0 except when $d_0=1$ for which the slope is $-1/\lc$.
Using equations \eqref{eq:exactPFStress} and \eqref{eq:gradex}, we obtain the displacement field for $U_c \leq U \leq\wc$ and $d_0 < 1$:
\begin{equation}
 \uP(x) =
 \left\lbrace
 \begin{tabular}{ll}
   $ - \displaystyle \frac{\sigc  (1-d_0)}{E} \left( \lc \int_{0}^{d_0} \frac{1}{ H(d,d_0) \omega(d)} \dint d - x - \ell \right) $ & if  $ -L/2  < x < -\ell$
\\
 $ - \displaystyle \frac{\sigc \lc (1-d_0)}{E}   \int_{\dP(x)}^{d_0} \frac{1}{ H(\tilde{d},d_0) \omega(\tilde{d})} \dint \tilde{d}$  & if  $ -\ell < x < 0$
 \\
 $ \displaystyle \frac{\sigc \lc (1-d_0)}{E}   \int_{\dP(x)}^{d_0} \frac{1}{ H(\tilde{d},d_0) \omega(\tilde{d})} \dint \tilde{d}$  & if   $ 0 < x < \ell$
\\
   $ \displaystyle \frac{\sigc  (1-d_0)}{E} \left( \lc \int_{0}^{d_0} \frac{1}{ H(d,d_0) \omega(d)} \dint d + x - \ell \right) $ & if  $ \ell < x < L/2$
 \end{tabular}
\right.
\label{eq:uExP}
\end{equation}
while for $d_0 = 1$:
\begin{equation}
 \uP(x) =
 \left\lbrace
 \begin{tabular}{rc}
$  \displaystyle -\frac{\wc}{2} $ & if  $ x < 0 $
 \\ &
  \\
$ \displaystyle \frac{\wc}{2} $ & if  $ x> 0. $
 \end{tabular}
\right.
\end{equation}

\section{Lip-field}
\label{appendix:lipfield}

The damage profile is
\begin{equation}
 \dL(x) =
  \left\lbrace
 \begin{tabular}{cc}
$   d_0 - \displaystyle \frac{|x|}{\lc} $ & if  $ |x| \leq d_0 \lc $
 \\ &
  \\
$ 0 $ & if  $ |x|  > d_0 \lc $
 \end{tabular}
\right.
\end{equation}
and the displacement field is for $U_c \leq U \leq\wc$ and $d_0 < 1$:
\begin{equation}
 \uL(x) =
 \left\lbrace
 \begin{tabular}{ll}
   $ - \displaystyle \frac{\sigc  (1-d_0^2)}{E}  \left[ \lc \left( d_0-\dL(x) + \frac{1}{\gamma} \left(\frac{1}{1-d_0^2} - \frac{1}{1-\dL^2(x)} \right)  \right) - x - d_0 \lc   \right] $ & if  $ -L/2  < x < - d_0 \lc$
\\
 $ - \displaystyle \frac{\sigc  (1-d_0^2)}{E} \lc \left[  d_0-\dL(x) + \frac{1}{\gamma} \left(\frac{1}{1-d_0^2} - \frac{1}{1-\dL^2(x)} \right)  \right] $  & if  $ -\ell < x < 0$
 \\
 $ \displaystyle \frac{\sigc  (1-d_0^2)}{E} \lc \left[  d_0-\dL(x) + \frac{1}{\gamma} \left(\frac{1}{1-d_0^2} - \frac{1}{1-\dL^2(x)} \right)  \right] $ & if   $ 0 < x < d_0 \lc$
\\
   $ \displaystyle \frac{\sigc  (1-d_0^2)}{E}  \left[ \lc \left( d_0-\dL(x) + \frac{1}{\gamma} \left(\frac{1}{1-d_0^2} - \frac{1}{1-\dL^2(x)} \right)  \right) + x - d_0 \lc   \right] $ & if  $ d_0 \lc < x < L/2$
 \end{tabular}
\right.
\label{eq:uExL}
\end{equation}
while for $d_0 = 1$:
\begin{equation}
 \uL(x) =
 \left\lbrace
 \begin{tabular}{rc}
$  \displaystyle -\frac{\wc}{2} $ & if  $ x < 0 $
 \\ &
  \\
$ \displaystyle \frac{\wc}{2} $ & if  $ x> 0.$
 \end{tabular}
\right.
\end{equation}

\bibliographystyle{unsrtnat}
\bibliography{references}  
\end{document}